\documentclass[journal]{IEEEtran}
\usepackage[absolute,overlay]{textpos}
\usepackage{accents}
\usepackage{amsfonts}
\usepackage{amssymb}
\usepackage{amsthm}
\usepackage{array}
\usepackage{textcomp}
\usepackage{stfloats}
\usepackage{url}
\usepackage{verbatim}
\usepackage{graphicx}
\usepackage{siunitx}

\usepackage{mathtools}
\usepackage{dsfont}
\usepackage{bm}
\usepackage{mathrsfs}
\usepackage{enumitem}
\usepackage{xpatch}
\usepackage{caption}
\usepackage{subcaption}
\usepackage{adjustbox}
\usepackage[ruled,norelsize, noline]{algorithm2e}
\usepackage{soul}
\usepackage{tikz}
\usepackage{tikzscale}
\usetikzlibrary{calc}
\usetikzlibrary{arrows, positioning}
\usetikzlibrary{shapes}
\usetikzlibrary{pgfplots.groupplots}
\usetikzlibrary{spy}
\usepackage{pgfplots}
\usepackage{scalefnt}
\usetikzlibrary{arrows.meta}
\usetikzlibrary{shapes.misc}
\usepackage[font=small]{caption}
\usepackage{pgfplots}
\usepgfplotslibrary{groupplots}
\usepackage{xcolor}
\usepackage[nolist]{acronym}

\usepackage{cite}
\usepackage{color}
\usepackage[hidelinks]{hyperref}                                                                                                      
\usepackage{orcidlink}	

\newacro{ATC-FWI}[ATC-FWI]{Adapt-Then-Combine Full Waveform Inversion}
\newacro{FWI}[FWI]{Full Waveform Inversion}
\newacro{NN}[NN]{Neural Network}
\newacro{ML}[ML]{Machine Learning}
\newacro{JSCC}[JSCC]{Joint Source and Channel Coding}
\newacro{SSCC}[SSCC]{Separate Source and Channel Coding}
\newacro{FL}[FL]{Federated Learning}
\newacro{OAC}[AirComp]{Over-the-Air Computation}
\newacro{ELBO}[ELBO]{Evidence Lower BOund}
\newacro{MILBO}[MILBO]{Mutual Information Lower BOund}
\newacro{dELBO}[dELBO]{Distributed Evidence Lower BOund}
\newacro{dMILBO}[dMILBO]{Distributed Mutual Information Lower BOund}
\newacro{AWGN}[AWGN]{Additive White Gaussian Noise}
\newacro{DNN}[DNN]{Deep Neural Network}
\newacro{SGD}[SGD]{Stochastic Gradient Descent}
\newacro{MSE}[MSE]{Mean Square Error}
\newacro{MAS}[MAS]{multi-agent system}  %{Multi Agent System}
\newacro{CNN}[CNN]{Convolutional Neural Network}
\newacro{NMSE}[NMSE]{Normalized Mean Square Error}
\newacro{DFC}[DFC]{Distributed Function Computation}
\newacro{PDE}[PDE]{Partial Differential Equation}
\newacro{KL}[KL]{Kullback-Leibler}

\newcommand{\s}{s}

\DeclareMathOperator*{\argmin}{argmin}
\DeclareMathOperator*{\argmax}{argmax}

\newcommand{\R}{\mathbb{R}}

\newcommand{\vsym}[1]{\ensuremath{\boldsymbol{#1}}}    % (fette) Vektoren oder Matrizen (mit Symbol als Argument)
\DeclareMathAlphabet\mathboldsf{OT1}{cmss}{bx}{n}

\newcommand{\T}{\mathsf{T}}
\newcommand{\rund}[1]{\left(#1\right)}

\newcommand{\obs}{\ensuremath{\mathrm{obs}}}
\newcommand{\syn}{\ensuremath{\mathrm{syn}}}
\newcommand{\iter}{\ensuremath{k}}
\newcommand{\itn}{\ensuremath{{[\iter]}}}
\newcommand{\itup}{\ensuremath{{[\iter+1]}}}

\newcommand{\drm}{\mathrm{d}}

\newcommand{\Nx}{{\ensuremath{N_{\mathrm{x}}}}}

\newcommand{\Nz}{{\ensuremath{N_{\mathrm{z}}}}}

\newcommand{\noRec}{{\ensuremath{N_{\mathrm{R}}}}}
\newcommand{\noSrc}{{\ensuremath{N_{\mathrm{S}}}}}

\newcommand{\idxRec}{r}
\newcommand{\idxSrc}{s}

\newcommand{\idxNeigh}{\ell}

\newcommand{\idxNeighNode}{{\idxNeigh\idxRec}}

\newcommand{\setNeighNode}{{\mathcal{N}_{\idxRec}}}

\newcommand{\neighIndexZero}{{\idxRec}}
\newcommand{\neighIndexOne}{{\idxRec\text{-}n_1}}
\newcommand{\neighIndexTwo}{{\idxRec\text{-}n_2}}
\newcommand{\neighIndexLast}{{{\idxRec\text{-}n_{N_r}}}}
\newcommand{\numberReceivers}{N_r}

\newcommand{\lowhat}[1]{\accentset{\smash{\raisebox{-2.1ex}{$\widehat{\phantom{#1}}$}}}{#1}}
\newcommand{\gradIndexZero}{\delta\vecModel_{{\idxRec}}}
\newcommand{\gradIndexOne}{\delta\vecModel_{{\idxRec\text{-}n}_{1}}}
\newcommand{\gradIndexLast}{\delta\vecModel_{{\idxRec\text{-}n}_{N_r}}}
\newcommand{\gradIndexEll}{\delta\vecModel_{{\ell}}}
\newcommand{\gradIndexIhat}{\lowhat{ \delta\vecModel}_{{\ell}}}
\newcommand{\gradIndexOnehat}{\lowhat{ \delta\vecModel}_{{\idxRec\text{-}n}_{1}}}
\newcommand{\gradIndexLasthat}{\lowhat{ \delta\vecModel}_{{{\idxRec\text{-}n}_{N_r}}}}

\newcommand{\gradSum}{\overline{\delta\vecModel}_{{\idxRec}}}
\newcommand{\gradSumEstimate}{\widehat{\overline{\delta\vecModel}}_{{\idxRec}}}

\newcommand{\veloIndexZero}{\widetilde{\vecModel}_{{\idxRec}}}
\newcommand{\veloIndexOne}{\widetilde{\vecModel}_{{\idxRec\text{-}n}_{1}}}
\newcommand{\veloIndexLast}{\widetilde{\vecModel}_{{\idxRec\text{-}n}_{N_r}}}
\newcommand{\veloIndexEll}{\widetilde{\vecModel}_{{\ell}}}
\newcommand{\veloIndexOneHat}{\widehat{\widetilde{\vecModel}}_{{\idxRec\text{-}n}_{1}}}
\newcommand{\veloIndexLastHat}{\widehat{\widetilde{\vecModel}}_{{\idxRec\text{-}n}_{N_r}}}
\newcommand{\veloIndexIhat}{\widehat{ \widetilde{\vecModel}}_{{\ell}}}
\newcommand{\veloIndexOnehat}{\widehat{ \widetilde{\vecModel}}_{{\idxRec\text{-}n}_{1}}}
\newcommand{\veloIndexLasthat}{\widehat{ \widetilde{\vecModel}}_{{\idxRec\text{-}n}_{N_r}}}

\newcommand{\veloSumi}{m_{r_i}}

\newcommand{\veloSum}{{{\vecModel}}_{{\idxRec}}}
\newcommand{\veloSumEstimate}{\widehat{{{\vecModel}}}_{{\idxRec}}}

\newcommand{\cost}{\ensuremath{\mathcal{L}}}

\newcommand{\wavefield}{\ensuremath{u}}

\newcommand{\source}{\ensuremath{f}}

\newcommand{\stepSize}{\ensuremath{\alpha}}

\newcommand{\model}{\ensuremath{m}}

\newcommand{\scaData}{d}

\newcommand{\vecModel}{\ensuremath{\vsym{\model}}}
\newcommand{\vecCoord}{\ensuremath{\vsym{x}}}

\newcommand*{\linewidthtikz}{1.0pt}
\newcommand*{\marksize}{2.0} %2.6
\newcommand*{\marksizeRed}{0.8}
\newcommand*{\marksizeBlue}{2.0}

\newlength\figH
\newlength\figW
\newlength\figWsmall
\newcommand{\legendfontsize}{\normalsize}
\newcommand{\ltickfontsize}{\tiny} % normalsize
\newcommand*{\labelaxisfontsize}{\small}

\newcommand{\markrepeat}{1}
\newcommand{\xminimum}{0}
\newcommand{\xmaximum}{0.7}
\newcommand{\yminimum}{-20}
\newcommand*{\ymaximum}{0}

\newcommand*{\nth}[2]{
    \foreach \xindx [count=\kcount from 0] in #1 {
        \ifnum\kcount=#2
            \xindx
        \fi
    }
}

\newcommand{\xlabel}{time$[dB]$}
\newcommand{\ylabel}{system mismatch $[dB]$}

\newboolean{xtickAutomatic}
\setboolean{xtickAutomatic}{true}
\newcommand{\xticklabelList}{0, 0.1, 0.3, 0.5}
\newboolean{ytickAutomatic}
\setboolean{ytickAutomatic}{true}
\newcommand{\yticklabelList}{0, 0.1, 0.3, 0.5}

\newboolean{xMinAutomatic}
\setboolean{xMinAutomatic}{true}
\newboolean{xMaxAutomatic}
\setboolean{xMaxAutomatic}{true}

\newboolean{yMinAutomatic}
\setboolean{yMinAutomatic}{true}
\newboolean{yMaxAutomatic}
\setboolean{yMaxAutomatic}{true}

\newboolean{LegendAutomatic}
\setboolean{LegendAutomatic}{true}

\newcommand{\legendtranspose}{ }
\newcommand{\legendlabela}{1}
\newcommand{\legendlabelb}{2}
\newcommand{\legendlabelc}{3}
\newcommand{\legendlabeld}{4}
\newcommand{\legendlabele}{5}
\newcommand{\legendlabelf}{6}

\ifCLASSINFOpdf
\else
\fi
\begin{document}
\title{Semantic Joint Source Channel Coding\\ for Distributed Subsurface Imaging\\ in Multi-Agent Systems}

\author{Maximilian H. V. Tillmann,~\IEEEmembership{Graduate Student Member,~IEEE,}
        Ban-Sok Shin,~\IEEEmembership{Member,~IEEE,}        
        Dmitriy~Shutin,~\IEEEmembership{Senior Member,~IEEE,}
        Armin Dekorsy,~\IEEEmembership{Senior Member,~IEEE}
        % <-this % stops a space
\thanks{
This work was supported in part by the Deutsche Forschungsgemeinschaft (DFG, German Research Foundation)  under grant 500260669 (SCIL), by the DFG under Germany's Excellence \mbox{Strategy – EXC-3036} The Martian Mindset, project number: 533607631, and by
the German Ministry of Research, Technology and Space (BMFTR) under project number 16KISK016 (Open6GGub).
}%
%This work was supported by the German Research Foundation (DFG) under grant 500260669 (SCIL).%
\thanks{Maximilian H. V. Tillmann and Armin Dekorsy are with the Department of Communications Engineering,
University of Bremen, 28359 Bremen, Germany (e-mail: tillmann{\scriptsize@}ant.uni-bremen.de; dekorsy{\scriptsize@}ant.uni-bremen.de).}%
\thanks{Ban-Sok Shin and Dmitriy Shutin are with the Institute of Communications and Navigation, German
Aerospace Center, 82234 Wessling, Germany (e-mail: ban-sok.shin{\scriptsize@}dlr.de; dmitriy.shutin{\scriptsize@}dlr.de).}%
%\thanks{M. Shell was with the Department of Electrical and Computer Engineering, Georgia Institute of Technology, Atlanta, GA, 30332 USA e-mail: (see http://www.michaelshell.org/contact.html).}% <-this % stops a space
}


%\markboth{Submit to IEEE Transactions on Signal and Information Processing over Networks}%

\begin{textblock*}{2\columnwidth}(1.9cm,26.8cm)
\centering
\footnotesize
DOI: 10.1109/TSIPN.2026.3719310 ~\copyright~2026 IEEE \url{https://ieeexplore.ieee.org/document/11636215}
\end{textblock*}



\maketitle

\begin{abstract}
Multi-agent systems (MASs) are a promising solution for autonomous exploration tasks in hazardous or remote environments. In such settings, communication among agents is essential to ensure collaborative task execution, yet conventional approaches treat exploration and communication as decoupled subsystems. This work presents an approach that tightly integrates semantic communication into the MAS exploration process, adapting the communication system to the exploration methodology to improve overall task performance. Specifically, we investigate the application of semantic joint source-channel coding (JSCC) with over-the-air computation (AirComp) for distributed function computation for the application of cooperative subsurface imaging using the adapt-then-combine full waveform inversion (ATC-FWI) algorithm. Our results demonstrate that semantic JSCC significantly outperforms classical digital communication and conventional JSCC approaches, especially in high-connectivity networks. Furthermore, incorporating side information at the receiving agent enhances communication efficiency and imaging accuracy, a feature previously unexplored in MAS-based exploration. We validate our approach through a use case inspired by subsurface anomaly detection, showing measurable improvements in imaging performance per agent. This work underscores the potential of semantic communication in distributed multi-agent exploration to improve overall exploration accuracy and efficiency.
\end{abstract}


% \fi
\IEEEpeerreviewmaketitle

\section{Introduction}
% Max. 13 pages

%\subsection{Motivation}

\begin{figure}[tb]
    \centering
    \includegraphics[width=0.5\textwidth]{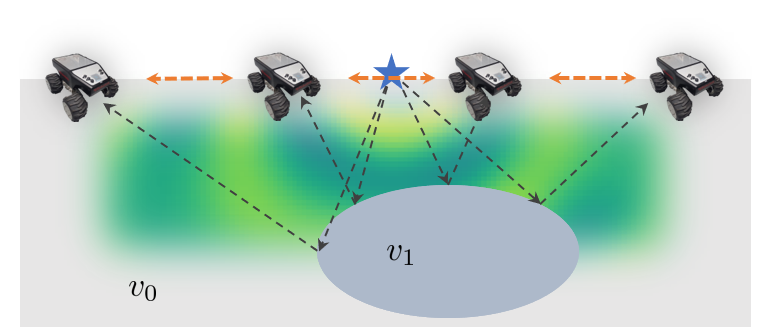}
    \caption{Illustration of the exploration scenario: A \ac{MAS} consisting of ground rovers reconstructs the velocity structure $(v_0, v_1)$ of a subsurface using a seismic source (blue) and seismic waves reflecting from an anomaly. The rovers are connected over wireless links (orange) that enable data exchange within the MAS. The communication among rovers is realized via semantic communication to adapt the communication system to the exploration algorithm.}
    \label{fig:scenario}
\end{figure}

\IEEEPARstart{M}{ulti-agent} systems have emerged as a crucial tool to address various tasks that cannot be handled reliably by humans such as inspection, surveying, monitoring, and exploration of hazardous or remote environments \cite{Zhang2020, Nathan2024}. For instance, \acp{MAS} have been proposed and investigated for the use in planetary exploration, gas and subsurface exploration \cite{Rabideau2025, Wiedemann2019, Shin2024}. A \ac{MAS} inherently requires communication among its agents to achieve the respective exploration goal such as reconstruction of a physical process or estimation of certain parameters. Usually, the information exchange among agents is realized by a wireless communication system and therefore, most works use standard communication schemes to achieve this, cf. \cite{Rizk2025, Zhang2025}. In this respect, exploration and communication are viewed as two systems that are designed separately although exploration and communication are often strongly entangled in a \ac{MAS}. To achieve enhanced exploration performance both communication and exploration need to be considered jointly. Therefore, in this work, we propose to integrate semantic communication into a distributed exploration algorithm such that the overall system's exploration performance is enhanced. The specific exploration task considered here is subsurface exploration by means of seismic imaging which is illustrated in Fig. \ref{fig:scenario}. 
% By doing so, the communication system is made aware of the exploration task. 

Semantic communication is a promising technology that enables task-specific communication beyond the Shannon limit of classical communication systems \cite{survey_gunduz2022,survey_yang2022,survey_wheeler2023,xu2023edge}.
Compared to classical communication systems semantic communication aims to transmit only the required ``meaning'' of the message for accurate task execution to save bandwidth \cite{survey_gunduz2022}. 
% Most research on semantic communication focuses on either task execution or data reconstruction \cite{survey_gunduz2022}. 
% In this work, we focus on semantic communication for task execution. 
Recently, modern machine learning techniques have been used to develop such semantic communication systems. 
Such a system becomes especially advantageous in distributed exploration scenarios, where a task needs to be solved by multiple agents. In this case, semantic communication makes it possible to only exchange information relevant for the exploration task between the agents instead of transmitting the complete data.

One envisioned application scenario of \acp{MAS} is the autonomous quantification of lunar caves in future exploration missions using seismic techniques \cite{Shin2024, Kalita2021, delaCroix2024}. 
Seismic methods typically require spatially distributed, simultaneous measurements. A \ac{MAS} with agents carrying seismic sensors thus perfectly exemplifies the use case of interest.
% Lunar caves and lava tubes are relevant for future space missions: Since they provide constant temperature and natural protection from radiation, they are promising sites to be used as habitats for space equipment or even astronauts. 
In this regard, the \ac{ATC-FWI} has been proposed, which enables subsurface imaging in a distributed fashion within a \ac{MAS} \cite{Shin2024}. In essence, ATC-FWI obtains subsurface images at each agent via data exchange among neighboring agents using distributed data processing. One fundamental operation in such distributed processing is averaging of exchanged data, which can be found in e.g., consensus averaging and diffusion-based schemes \cite{Sayed2013, Shi2015}. From a semantic communication perspective the distributed averaging process of several input data at the receiver belongs to the problem of \emph{distributed function computation} \cite{ma2011some,abari2016over}. It is at this point where we apply semantic communication techniques to appropriately encode the data of multiple transmitters (transmitting agents) to perform \ac{DFC} for the averaging step at the decoder (receiving agent). We employ a probabilistic model for the encoders and the decoder of the proposed semantic communication system, where we assume that the decoder is modeled with independent Gaussian distributions for every vector element.
In this work, we propose the use of semantic communication in a distributed subsurface imaging method for a \ac{MAS} to obtain reliable imaging results for noisy inter-agent communication links. To the best of our knowledge, this is the first work to employ semantic communication for distributed exploration in a \ac{MAS}. We summarize the main contributions as follows:

\begin{itemize}
    \item \textbf{Semantic DFC:} Since the agents reconstruct the same subsurface from different positions, a receiving agent has correlated side information available. We exploit this fact by formulating the semantic \ac{DFC} problem as a maximization of the mutual information given the side information, i.e., the received symbols should be maximally informative about the semantic variable given the side information. Since maximizing the mutual information directly is infeasible (cf. \cite{vincent2010stacked, edgar2023semantic}), we derive a lower bound of the conditional mutual information, which can be maximized using end-to-end training. In previous works on semantic communication, either no side information was considered \cite{edgar2023semantic}, or an evidence lower bound was used when side information was present \cite{Neural_DSC}.
    \item \textbf{Integration to MAS:} We apply our proposed semantic \ac{DFC} to the data exchange procedure within the \ac{ATC-FWI}. We employ \ac{OAC} for semantic \ac{DFC} to use channel resources between agents more efficiently. To this end, we train the encoders and the decoder for \ac{OAC} in an end-to-end manner using \acp{DNN}. This allows us to express the decoder as a nomographic function, which is necessary to enable addition of the encoded symbols over the communication channel.
    \item \textbf{Results:} In simulations, we show that the proposed method can improve bandwidth efficiency for inter-agent connections corrupted by white Gaussian noise in \ac{ATC-FWI} compared to state-of-the-art \ac{JSCC} methods. Our proposed approach of semantic \ac{DFC} outperforms state-of-the-art \ac{JSCC} methods, especially in scenarios, where many neighboring agents exchange data with each other. In particular, compared to a classical digital transmission our semantic \ac{DFC} shows significantly improved communication performance and thus, better imaging performance for \ac{ATC-FWI}.   
    %It should be noted that our proposed semantic \ac{DFC} is also applicable to other channel models such as fading channels. \todo{@Max please check}
\end{itemize}

\section{Distributed Subsurface Imaging for Multi-Agent Systems}
In the following section, we briefly describe the \ac{FWI} as the classical, centralized imaging scheme from geophysics. Afterwards, we explain how \ac{FWI} can be adapted to a \ac{MAS} for distributed imaging, when each agent obtains a global subsurface image locally. This leads to the \ac{ATC-FWI} algorithm which requires explicit data exchange among neighboring agents and thus, suitable communication.

\subsection{Background}
To address the exploration problem in terms of subsurface imaging we consider a technique called \ac{FWI}. In particular, \ac{FWI} is a geophysical imaging technique that enables reconstruction of subsurface structures by exploiting wave physics \cite{Virieux2009, Fichtner2009}. Despite being rooted in seismic subsurface reconstruction, \ac{FWI} is applied to various other modalities such as ultrasound, ground penetrating radar and electrical resistivity tomography. In its traditional form, FWI is a centralized scheme that requires all measurement data to be allocated at a central entity that performs imaging. Hence, for application in a \ac{MAS} it needs to be adapted to enable imaging at each agent via exchange of data within the network. In our previous work, we proposed the \ac{ATC-FWI} which enables distributed imaging within a \ac{MAS} \cite{Shin2024}. 
However, for simplicity \ac{ATC-FWI} has been considered with error-free communication among the agents in the previous work, which cannot be achieved with noisy communication channels.
%However, \ac{ATC-FWI} has been considered error-free communication among the agents, i.e., without considering any communication aspects. 
Other distributed or decentralized schemes for subsurface imaging have been proposed in the literature, cf. \cite{Li2020, Wang2021}. However, in these works communication aspects are considered separately from the imaging problem. That is, the communication is realized by using standard wireless communication systems and an adapted design of the communication system to the subsurface imaging is not considered. 

\subsection{Centralized Full Waveform Inversion}
In its ordinary formulation, FWI aims at minimizing a least-squares cost functional of a data residual with respect to a subsurface parameter $\model(\vecCoord)$ that varies over space coordinate $\vecCoord$. The residual is evaluated between observed measurements $\scaData^\obs(t)$ and predicted or synthesized measurements $\scaData^\syn(t,\model)$ over time $t$. The synthesized measurements $\scaData^\syn(t,\model)$ are generated by solving the acoustic wave equation and sampling the obtained wavefield $u_\idxSrc(\vecCoord,t)$ at the receiver positions via a sampling operator $S_{\idxSrc,\idxRec}$ for a receiver~$\idxRec$ and the respective source~$\idxSrc$. The fact that the minimization is coupled to the acoustic wave equation can be formulated as a constraint in the resulting optimization problem:
\begin{subequations}
\begin{align}
	\underset{\model}{\text{argmin}} \, \mathcal{L}(\model) &= \frac{1}{2} \sum_{\idxSrc,\idxRec} \int_0^\tau [ \underbrace{S_{\idxSrc,\idxRec} u_\idxSrc(\vecCoord,t)}_{\scaData_{\idxSrc,\idxRec}^\syn(t, \model)} - \scaData_{\idxSrc,\idxRec}^\obs(t) ]^2 \, \drm t
    \label{eq:fwi_cost}\\
    \mathrm{s.t.} &\quad \frac{1}{\model^2(\vecCoord)} \frac{\partial^2 u_\idxSrc(\vecCoord, t)}{\partial t^2} - \frac{\partial^2 u_\idxSrc(\vecCoord, t)}{\partial \vecCoord^2} = \source_\idxSrc(\vecCoord, t).
    \label{eq:awe}
\end{align}
\label{eq:fwi_problem}
\end{subequations}
where $\sum_{\idxSrc,\idxRec}:=\sum_{\idxSrc=1}^\noSrc \sum_{\idxRec=1}^\noRec$ and $\noSrc,\noRec$ is the number of seismic sources and receivers, respectively. In our case, the subsurface parameter $\model(\vecCoord)$ is a function over space and represents the spatial distribution of the $P$-wave velocity. As can be seen, FWI essentially aims at solving an optimization problem that is constrained by a \ac{PDE}. In the wave equation~\eqref{eq:awe}, $\source_\idxSrc(\vecCoord, t)$ is the seismic source specific to source~$\idxSrc$. Solving \eqref{eq:awe} for a subsurface model $\model(\vecCoord)$ results in the wavefield $\wavefield_\idxSrc(\vecCoord, t)$ which is then sampled at the receiver positions $\{\vecCoord_\idxRec\}_{\idxRec=1}^{\noRec}$ to give $\scaData^\syn_{\idxSrc,\idxRec}(t,\model)$.

%To evaluate the data residual in \eqref{eq:fwi_cost} seismic data $\scaData^\syn_{\idxSrc,\idxRec}$ need to be synthesized for each receiver~$\idxRec$ and source~$\idxSrc$. To this end, the acoustic wave equation needs to be solved which is given in the following form:
%\begin{equation}
%    \frac{1}{\model^2(\vecCoord)} \frac{\partial^2 u_\idxSrc(\vecCoord, t)}{\partial t^2} - \frac{\partial^2 u_\idxSrc(\vecCoord, t)}{\partial \vecCoord^2} = \source_\idxSrc(\vecCoord, t).
%    \label{eq:awe}
%\end{equation}
%with initial conditions
% \begin{equation}
%	\wavefield_\idxSrc(\vecCoord, 0) = 0, \quad\frac{\partial %\wavefield_\idxSrc(\vecCoord, 0)}{\partial t} = 0.
% \end{equation}

Problem~\eqref{eq:fwi_problem} is nonlinear in $\model(\vecCoord)$ and therefore requires gradient-based optimization to converge to a suitable subsurface model. Commonly used methods include gradient-descent, nonlinear conjugate gradient or L-BFGS. To this end, the gradient of $\cost(\model)$ needs to be calculated. Note that $\cost(\model)$ is constrained by a  \ac{PDE} which complicates direct calculation of the gradient. Typically, the so-called \emph{adjoint state method} is used to compute the gradient of cost functionals that are constrained by a PDE, cf. \cite{Plessix2006}. Applying the adjoint state method to \eqref{eq:fwi_cost} together with~\eqref{eq:awe} results in the following gradient computation \cite{Shin2024}:
\begin{equation}
	\frac{\drm\cost(\model)}{\drm \model} = - \frac{2}{\model(\vecCoord)^3} \sum_{\idxSrc=1}^\noSrc \int_0^\tau q_\idxSrc(\vecCoord, \tau-t) \frac{\partial^2 u_\idxSrc(\vecCoord, t)}{\partial t^2} \drm t.
	\label{eq:fwi_gradient}
\end{equation}
Here, the function $q_\idxSrc(\cdot, \cdot)$ is a so-called adjoint wavefield, which is obtained by solving the following \ac{PDE} which is an adjoint wave equation:
\begin{align}
	\rund{\frac{1}{\model(\vecCoord)^2} \frac{\partial^2 }{\partial t ^2}-\frac{\partial^2 }{\partial \vecCoord ^2}} q_\idxSrc(\vecCoord, t) &= \sum_{\idxRec=1}^\noRec S_{\idxSrc,\idxRec} (d^\syn_{\idxSrc,\idxRec}(\tau-t)\nonumber\\&-d^\obs_{\idxSrc,\idxRec}(\tau-t)).
	\label{eq:fwi_adjoint_field}
\end{align}
% with initial conditions $q_\idxSrc(0) = 0, \partial q_\idxSrc(0)/\partial t = 0$. 
Note that to solve~\eqref{eq:fwi_adjoint_field}, the data residuals $(d^\syn_{\idxSrc,\idxRec}(\tau-t)-d^\obs_{\idxSrc,\idxRec}(\tau-t))$ are now used as a source signal that is reversed in time and injected at the receiver positions by a sampling operator $S_{\idxSrc,\idxRec}$. After obtaining the adjoint wavefield, $q_\idxSrc(\vecCoord, t)$ is again reversed in time and correlated with the second time derivative of the wavefield $u_\idxSrc(\vecCoord, t)$ following \eqref{eq:fwi_gradient}.

To solve PDEs in \eqref{eq:awe} and \eqref{eq:fwi_adjoint_field} we require a numerical solver on a discrete space and time grid. In our work, we use the Devito package that provides highly efficient finite difference solvers for PDEs \cite{Louboutin2019}. By discretizing the considered 2D computational domain into a $\Nx\times\Nz$ grid we obtain vector quantities for the subsurface model $\model(\vecCoord)$ and the gradient function $\drm\cost(\model)/\drm\model$. Therefore, we replace these functions by their discrete vector description, i.e., 
\begin{equation}
    \model(\vecCoord) \rightarrow \vecModel(\vecCoord)\in\R^{\Nx\Nz}, \quad \drm\cost(\model)/\drm\model \rightarrow \delta\vecModel\in\R^{\Nx\Nz}.
\end{equation}
Given a gradient by~\eqref{eq:fwi_gradient}, we can now minimize $\cost(\model)$ iteratively using gradient descent. This results in the following update procedure for the estimated subsurface model:
\begin{equation}
    \vecModel^{\itup} = \vecModel^{\itn} - \stepSize^{\itn} \delta\vecModel^{\itn}
\end{equation}
The parameter $\stepSize^{\itn}>0$ is a step size that can be adapted over the iterations e.g. by a simple exponential decay or a line search method.

\subsection{Adapt-Then-Combine Full Waveform Inversion (ATC-FWI)}
FWI in its classical setting is centralized algorithm, i.e., it requires all measurement data $\scaData^\obs_{\idxSrc,\idxRec}$ to be available at a central entity to perform the inversion procedure. However, in the \ac{MAS} no central entity is available that collects all measurement data from the agents. Instead, we demand a computation within the \ac{MAS} by exploiting agent-to-agent communication.

The objective of a distributed imaging scheme is to obtain an image locally at each agent/receiver via cooperation with neighboring agents, i.e., those that are in the communication range of the respective agent. To this end, the ATC-FWI has been proposed in \cite{Shin2024}. In the following, we give a short summary of the algorithm.

Let us first introduce assumptions about the network topology of the MAS. We assume a network of $\noRec$ agents, where each agent is equipped with a geophone or a seismic sensor. Each agent~$\idxRec$ can communicate with neighboring agents~$\idxNeigh$ that are contained in a neighborhood set $\setNeighNode \subseteq\{1,\ldots,\noRec\}$ and vice versa. We assume that the neighborhood set $\setNeighNode$ contains the agent~$\idxRec$ itself. Furthermore, the network graph is undirected and connected, i.e., each agent can be reached by any other agent in the network over multiple hops. 

The first step in deriving ATC-FWI is to represent the global cost in \eqref{eq:fwi_cost} over all $\noRec$ agents in the network such that 
\begin{equation}
    \mathcal{L} (\model) = \sum_{\idxRec=1}^\noRec \mathcal{L}_\idxRec (\model) \label{eq:global_cost},
\end{equation}
with each agent's local cost given now as
\begin{equation}
    \mathcal{L}_\idxRec (\model) = \frac{1}{2} \sum_{\idxSrc=1}^\noSrc \int_0^\tau \rund{\scaData_{\idxSrc,\idxRec}^\syn(t, \model) - \scaData_{\idxSrc,\idxRec}^\obs(t)}^2 \, dt.
    \label{eq:local_cost}
\end{equation}
Note that each local cost $\mathcal{L}_\idxRec (\model)$ depends only on locally observed measurement data $\scaData_{\idxSrc,\idxRec}^\obs(t)$. Based on its local cost each agent~$\idxRec$ is able to compute a gradient w.r.t. the subsurface model $\model$ in the same manner as in centralized FWI, however, with local measurement data $\scaData_{\idxSrc,\idxRec}^\obs(t)$. To ensure that global subsurface information is available to all agents in the network, we introduce~$\noRec$ different (local) subsurface models which are then combined using diffusion strategies~\cite{Sayed2013}. Specifically, we assign individual subsurface models $\model_\idxRec(\vecCoord)$ to each agent. Then each agent solves the wave equation~$\eqref{eq:awe}$ and the adjoint equation~\eqref{eq:fwi_adjoint_field} using its own subsurface model~$\model_\idxRec(\vecCoord)$ to obtain the wavefield $u_{\idxSrc, \idxRec}(\vecCoord,t)$ and the adjoint wavefield $q_{\idxSrc, \idxRec}(\vecCoord,t)$ according to
\begin{align}
    \frac{1}{\model_\idxRec(\vecCoord)^2} \frac{\partial^2 u_{\idxSrc,\idxRec}(\vecCoord, t)}{\partial t^2} - \frac{\partial^2 u_{\idxSrc,\idxRec}(\vecCoord, t)}{\partial \vecCoord^2}&= \source_\idxSrc(\vecCoord, t) \\
    \rund{\frac{1}{\model_\idxRec(\vecCoord)^2} \frac{\partial^2 }{\partial t ^2}-\frac{\partial^2 }{\partial \vecCoord ^2}} q_\idxSrc(\vecCoord, t) &= S_{\idxSrc,\idxRec}^\T (d^\syn_{\idxSrc,\idxRec}(\tau-t)\nonumber\\&-d^\obs_{\idxSrc,\idxRec}(\tau-t)),
\end{align}
followed by a local gradient computation $\delta\model_\idxRec(\vecCoord)$ via correlation:
\begin{equation}
    \delta\model_\idxRec(\vecCoord) = - \frac{2}{\model(\vecCoord)^3} \sum_{\idxSrc=1}^\noSrc \int_0^\tau q_{\idxSrc,\idxRec}(\vecCoord, \tau-t) \frac{\partial^2 u_{\idxSrc,\idxRec}(\vecCoord, t)}{\partial t^2} \drm t
\end{equation}
Now, each agent has an individual subsurface model~$\model_\idxRec(\vecCoord)$ and a local gradient $\delta\model_\idxRec(\vecCoord)$ available. To enable distributed imaging such that each agent's model approaches the centralized FWI result, we apply the adapt-then-combine (ATC) technique from \cite{Sayed2013}. Assuming again discretization of the spatial domain using finite differences we represent gradients and subsurface models via vectors. Applying ATC results in the following computations per agent~$\idxRec$:
\begin{subequations}
    \label{eq:atcfwi}
    \begin{align}
        \widetilde{\vecModel}_\idxRec^{\itup} &= \vecModel_\idxRec^{[\iter]} + \stepSize^{[\iter]} \sum_{\idxNeigh\in\setNeighNode} b_{\idxNeighNode} \delta\vecModel_\idxNeigh^{[\iter]} \label{eq:atcfwi_a}\\
        \vecModel_\idxRec^{\itup} &= \sum_{\idxNeigh\in\setNeighNode} a_{\idxNeighNode} \widetilde{\vecModel}_\idxNeigh^{\itup}\label{eq:atcfwi_b}
    \end{align}    
\end{subequations}
The coefficients $a_{\idxNeighNode}, b_{\idxNeighNode}$ need to be chosen appropriately such that the centralized FWI result can be approached by all agents, cf. \cite{Sayed2013}. In the simplest case, one can select these coefficients to be $1/|\setNeighNode|$, where $|\cdot|$ denotes cardinality of the neighborhood set. This results in a uniform averaging of neighboring data. 

From~\eqref{eq:atcfwi} we see that in each iteration~$\iter$ each agent~$\idxRec$ needs to exchange its gradient $\delta\vecModel_\idxRec$ and its intermediate subsurface model $\widetilde{\vecModel}_\idxRec$ with its neighbors~$\idxNeigh\in\setNeighNode$. In particular, the fusion process of neighboring data in ATC is given by an averaging process. It is here, where we will integrate semantic communications to exploit the averaging structure in the data exchange among the agents.

% \section{Semantic Communication System Model and Problem Formulation}
\section{Semantic Distributed Function Computation for ATC-FWI}

First, a background of semantic communication for \acp{MAS} is provided.
In what follows, we discuss the semantic communication approach towards computation of ATC over the network.
In particular, we will outline the design of corresponding encoder-decoder structures for computing the averaging operations in~\eqref{eq:atcfwi_a} and~\eqref{eq:atcfwi_b} over a wireless communication channel.
We already note here that, in order to design the proposed semantic communication system with probabilistic encoder and decoder models, we simplify the decoder distribution by modeling it with independent Gaussian distributions to efficiently solve the formulated optimization problem. The investigation of alternative decoder distribution types is left for future work.

%We already note here that, in order to design the proposed semantic communication system with probabilistic encoder and decoder models, we assume that the decoder is modeled using independent Gaussian distributions for each vector element.

\subsection{Background}

For semantic communication we want to compress and channel code the data at the transmitter such that we can reconstruct the semantic information at the receiver. To this end, deep learning techniques are used to learn an end-to-end encoder decoder system for \ac{JSCC} \cite{survey_gunduz2022,xu2023deep}.
Semantic communication systems become increasingly important in distributed settings with multiple transmitters and one or more receivers \cite{JSCC_survey}, quite typical for \ac{MAS} applications. For example, in
\cite{edgar2023semantic} and \cite{razlighi2024cooperative} semantic communication was investigated for the case when multiple transmitters with different observations need to cooperatively solve a classification task. In \cite{Distributed_Image_Transmission}, the authors investigate the case where one receiver that obtains two correlated images from two transmitters has to reconstruct both images.

The application of semantic communication to \ac{ATC-FWI} requires consideration of two key aspects: side information at the decoder, i.e., the receiving agent and \ac{DFC}. 
Side information should be incorporated into the semantic communication system for \ac{ATC-FWI}, as the data available at each agent are correlated and can therefore assist the decoding process, and \ac{DFC} should be considered since \ac{ATC-FWI} requires the computation of a function of the exchanged data, which is here the weighted average.
In this respect, the use of side information at the receiver for communication of images, was investigated in \cite{Neural_DSC} and \cite{Mital_2023_WACV}. In \cite{Neural_DSC}, it was shown for image transmission that the compression level for the case when side information is only available at the decoder can nearly achieve the same compression level when side information is available at both the encoder and the decoder. On the other hand, \ac{DFC} was studied in \cite{distrib_source_coding_FL}, with its variation with \ac{OAC} investigated in \cite{FL_OAC, OAC_channel_SNR}, and deep learning based \ac{OAC} in \cite{Deep_OAC, OAC_semantic_multi_user, Distributed_semantic_oac_sensing}. 
Furthermore, \ac{OAC} was used in \cite{agrawal2024distributed} for distributed convex optimization.
\ac{OAC} aims at using channel resources more efficiently by letting users transmit at the same time and frequency. In this way, a function can be evaluated directly on the wireless channel as a sum of the transmitted symbols by exploiting the superposition property of electromagnetic waves  \cite{OAC_survey1,OAC_survey2}. 
It was shown that despite advantages, \ac{OAC} mainly has two drawbacks: (i) 
the limitation that, with encoders $\bm{\varphi}_i$ and decoder $\bm{\phi}$, the transmitted symbols $\bm{\varphi}_i(\bm{s}_i)$ are required to be decodable after being added over the wireless communication channel, i.e., that they form a  so-called nomographic function $\bm{f}(\bm{s}_1,\dots,\bm{s}_N)=\bm{\phi}(\sum_{i=1}^N \bm{\varphi}_i(\bm{s}_i))$ and (ii) the requirement of accurate synchronization \cite{OAC_survey1,OAC_survey2}. 
A nomographic function is required since \ac{OAC} allows simultaneous multi-device transmission, resulting in a received signal that is the superposition of the individual carrier signal amplitudes \cite{OAC_survey1}.

\subsection{Notation}
In the following, we denote the receiving agent by the index $r$. We define  $\numberReceivers \triangleq |\setNeighNode|-1$ to count the number of neighbors of the receiving agent $r$, excluding the agent itself, as the neighborhood $\setNeighNode$ includes (by definition) the receiving agent $r$ as well.
In order to be able to denote the neighbors of any agent $r$ generally, we will use the following indices to address the elements of the neighbor set $\setNeighNode$: The agent $r$ itself will be denoted by $\neighIndexZero$, and all neighbors of agent $r$ will be denoted by $\neighIndexOne,\neighIndexTwo,\dots,\neighIndexLast$.
Furthermore, we omit the algorithm iteration index $k$ of the \ac{ATC-FWI} to simplify the notation.

\subsection{Semantic Communication System Model}

For each agent to compute \eqref{eq:atcfwi_a}, its neighbors have to share the velocity gradients $\gradIndexEll$. Likewise, for each agent to compute \eqref{eq:atcfwi_b}, its neighbors have to share the velocity models $\veloIndexEll$.
Our semantic communication system model for the exchange of the velocity gradients $\gradIndexEll$ is shown in Fig.~\ref{fig:system_model_example_gradient}, and for the exchange of the velocity models $\veloIndexEll$ it is shown in Fig.~\ref{fig:system_model_example_velovity}.

Although the velocity gradients $\gradIndexEll$  and the velocity models $\veloIndexEll$ are different quantities, the concept of semantic communication will be applied to each of them similarly.
Therefore, to avoid redundancy, the semantic communication system model and optimization problem will only be shown for the exchange of the velocity models $\veloIndexEll$ in the following.

%As semantic communication is applied analogously in both cases, in the following the concept will only be shown for the exchange of the velocity model $\veloIndexEll$.

We define a semantic source as the joint probability distribution $p(\veloIndexZero,\veloIndexOne,\dots,\veloIndexLast,\veloSum)$ of the observations $\veloIndexZero,\veloIndexOne,\dots,\veloIndexLast$ and the semantic variable $\veloSum$. The semantic variable in this case is defined as $\veloSum = \sum_{\idxNeigh\in\setNeighNode} a_{\ell r } \veloIndexEll$.
%-The goal is to reconstruct the semantic variable $\hat{\veloSum}$ at the receiver given the observations $\veloIndexZero,\dots,\bm{\s}_N$ and semantic variable $\veloSum$ from the distribution of the semantic source $p(\veloIndexZero,\dots,\bm{\s}_N,\veloSum)$.
The goal at the receiver is to reconstruct the semantic variable $\veloSum$, which together with the spatially distributed observations $\veloIndexZero,\veloIndexOne,\dots,\veloIndexLast$ follows the joint distribution of the semantic source $p(\veloIndexZero,\veloIndexOne,\dots,\veloIndexLast,\veloSum)$.
%-The goal is to reconstruct the semantic variable $\hat{\veloSum}$ at the receiver given the observations $\veloIndexZero,\dots,\bm{\s}_N$ with distribution $p(\veloIndexZero,\dots,\bm{\s}_N)$. With $p(\veloIndexZero,\dots,\bm{\s}_N,\veloSum) = p(\veloSum|\veloIndexZero,\dots,\bm{\s}_N)$$p(\veloIndexZero,\dots,\bm{\s}_N)$ the distribution of the semantic source is trying to be achieved. 

\begin{figure}[t!]
\centering
    % Resize to width=\textwidth
    \resizebox{0.5\textwidth}{!}{
    \includegraphics{tikz/block_diagram/system_model_example_gradient.tikz}
    }
    \caption{
    Block diagram of the multi-agent semantic DFC system for agent $r$ receiving the velocity gradients $\gradIndexEll$ from its neighbors. 
    The goal at the decoder is to directly reconstruct the semantic variable $\gradSumEstimate$, which is here the weighted average of the velocity gradients from the neighboring agents.
    }
    \label{fig:system_model_example_gradient}
\end{figure}

\begin{figure}[t!]
\centering
    % Resize to width=\textwidth
    \resizebox{0.5\textwidth}{!}{
    \includegraphics{tikz/block_diagram/system_model_example_velocity.tikz}
    }
    \caption{
    Block diagram of the multi-agent semantic DFC system for agent $r$ receiving the velocity models $\veloIndexEll$ from its neighbors. 
    The goal at the decoder is to directly reconstruct the semantic variable $\veloSumEstimate$, which is here the weighted average of the velocity models from the neighboring agents.
    }
    \label{fig:system_model_example_velovity}
\end{figure}

%Naturally, this ``local'' information should not be communicated since the agent is already aware about it.
%Instead, we will treat this information differently and utilize it to properly ``inform'' the decoder at the receiver.
%We will therefore refer to this data as \emph{side information} $\veloIndexZero$ in the following, using the subscript index $0$ to distinguish it from the other semantic sources.

%The observation $\veloIndexZero$ is already available at the decoder of the receiver, e.g., in the form of the local gradient or the local subsurface model of the receiving agent. We will therefore refer to this data as \emph {side information} for the decoding process.

The ``local'' observation $\veloIndexZero$ should not be communicated since the agent $r$ is already aware of it.
Instead, we will treat this information differently and utilize it to properly ``inform'' the decoder at the receiver.
We will therefore refer to the observation $\veloIndexZero$ as \emph {side information} for the decoding process.
The encoders $p_{{\bm{\theta}_\idxNeigh}}(\bm{c}_\idxNeigh|\bm{\s}_\idxNeigh)$ at the transmitting agents are modeled as \acp{NN} with parameters ${\bm{\theta}_\idxNeigh}$ to encode the observation data $\bm{\s}_\idxNeigh$ into symbols $\bm{c}_1,\dots,\bm{c}_{\numberReceivers}$, which are transmitted over a wireless channel $p(\bm{y}|\bm{c}_1,\dots,\bm{c}_{\numberReceivers})$, where $\bm{y}$ is a vector containing all received symbols. 
In our simulations in Section \ref{sec:simulation_results}, we focus on the simple case of an \ac{AWGN} channel. However, we note here that our probabilistic channel model is general, with the only requirement that $p(\bm{y}|\bm{c}_1,\dots,\bm{c}_{\numberReceivers})$ needs to be known. Hence, more complex channel models can be directly integrated into our framework.
%\todo{add info on AWGN for channel model, but approach works also for other channel models.}
To model the semantic decoder we use a variational approximation $q_{\bm{\phi}}(\veloSum|\bm{y},\veloIndexZero)$ with \ac{NN} parameters ${\bm{\phi}}$, which outputs the distribution of the semantic variable given the received symbols $\bm{y}$ and the side information $\veloIndexZero$. 
Finally, to get an estimate of the semantic variable $\veloSumEstimate$, we select the most likely $\veloSum$ from the posterior distribution $q_{\bm{\phi}}(\veloSum|\bm{y},\veloIndexZero)$.
The statistical dependencies among all random variables are summarized in Fig.~\ref{fig:markov_chain}.

% === SEMCOM SYSTEM MODEL ===== %
% \begin{figure}[t!]
% \centering
%     % Resize to width=\textwidth
%     \resizebox{0.5\textwidth}{!}{
%     \includegraphics{tikz/block_diagram/system_model.tikz}
%     }
%     \caption{Block diagram of multi-agent semantic DFC system with side information at the receiving agent $r$. The receiving agents gets the index $\idxNeigh=0$, and the neighboring agents are indexed from $1$ to $\numberReceivers$, with $\numberReceivers = |\setNeighNode|- 1$. Here, the observations $\bm{\s}_\idxNeigh$ are for the two cases the velocity estimates $\widetilde{\vecModel}_\idxNeigh$ or the gradients $\delta\vecModel_\idxNeigh$, and the semantic variable $\veloSum$ are the sums $\sum_{\idxNeigh\in\setNeighNode} a_{\idxNeighNode} \widetilde{\vecModel}_\idxNeigh$ or $\sum_{\idxNeigh\in\setNeighNode} b_{\idxNeighNode} \delta\vecModel_\idxNeigh$, respectively.
%      }
%     \label{fig:system_model}
% \end{figure}

\begin{figure}[t!]
\centering
    % Resize to width=\textwidth
    \resizebox{0.45\textwidth}{!}{
    \includegraphics{tikz/block_diagram/markov_chain_velo.tikz}
    }
    \caption{The statistical dependencies of the semantic variable $\veloSum$, the observations $\veloIndexEll$, the transmitted symbols $\bm{c}_\idxNeigh$, the received symbols $\bm{y}$, and the estimated sematic variable $\veloSumEstimate$.}
    \label{fig:markov_chain}
\end{figure}

% \subsection{Notation}
% $I(\cdot)$ is mutual information, $H(\cdot)$ is an entropy, $H(p(\cdot),q_{\bm{\phi}}(\cdot))$ is the cross entropy between distributions $p$ and $q$, and $E(\cdot)$ is an expected value. Throughout this paper we assume continuous probability distributions, unless stated otherwise.

\subsection{Semantic Communication Problem Formulation}
Our objective is to reconstruct the semantic variable $\veloSum$ at the decoder of each receiving agent.
We formulate the optimization problem of the semantic communication system such that the received symbols $\bm{y}$ are maximally informative about the semantic variable $\veloSum$, given the side information $\veloIndexZero$. 
This means that we want to find the parameters $\bm{\theta}_\idxNeigh$ for the encoders $p_{{\bm{\theta}_\idxNeigh}}(\bm{c}_\idxNeigh|\bm{\s}_\idxNeigh)$ that maximize the conditional mutual information $I(\veloSum;\bm{y}|\veloIndexZero)$.
To have a realistic communication scenario, we limit the rate of our communication channel by a limited number of channel uses $N_\text{ch}$ and an average power constraint per channel.
This gives us the optimization problem for receiving agent $\idxRec$ of
%$p_{{\bm{\theta}_\idxNeigh}}(\bm{c}_\idxNeigh|\bm{\s}_\idxNeigh) \ \text{for}\, i=1,\dots, \numberReceivers $
\begin{align} \label{InfoMax}
    &\argmax_{\bm{\theta}_1,\dots,\bm{\theta}_{\numberReceivers}}  I(\veloSum;\bm{y} |\veloIndexZero) \\
    & \text{s.t.} \ \ \bm{y}  \in \mathbb{R}^{\numberReceivers\cdot N_{\mathrm{ch}}} , \, P_c \leq 1 , \nonumber %\,\text{dB} \ \text{(limited signal power)} ,\nonumber 
\end{align}
where $\numberReceivers$ is the number of transmitters (i.e., the neighboring agents), and $P_c$ is the average signal power available at the decoder per channel use, which is defined differently depending on the particular channel model used (see Section \ref{sec:channel_models} for the definitions).

%See Section \ref{sec:channel_models} for the definition of $P_c$ for different channel models.

The definition of the conditional mutual information \mbox{between} $\veloSum$ and $\bm{y}$ given $\veloIndexZero$ is given as \cite{cover1999informationTheory}
%\begin{align}
% I(\veloSum;\bm{y} |\veloIndexZero) & = E_{p(\veloSum,\bm{y} ,\veloIndexZero)}\left[ \log\left( \frac{p(\veloSum,\bm{y} |\veloIndexZero)}{p(\veloSum|\veloIndexZero) p(\bm{y} |\veloIndexZero)} \right) \right] ,
%\end{align}
%which, with the chain rule for conditional probabilities can be rewritten to
\begin{equation}
    I(\veloSum;\bm{y} |\veloIndexZero)  =  E_{p(\veloSum,\bm{y} ,\veloIndexZero)}\left[ \log\left( \frac{p(\veloSum|\bm{y} ,\veloIndexZero)}{p(\veloSum|\veloIndexZero)} \right) \right] , \label{chain_rule_cond_pdf}
\end{equation}
where $\log$ denotes the natural logarithm.
 %Calculation of \eqref{chain_rule_cond_pdf} with the chain rules:
%\begin{align}
%    p(\veloSum,\bm{y} ,\veloIndexZero) &=  p(\veloSum|\bm{y} ,\veloIndexZero)p(\bm{y} %|\veloIndexZero) p(\veloIndexZero) \\
%                                & = p(\veloSum,\bm{y} |\veloIndexZero)p(\veloIndexZero)\\
%                        \iff \frac{p(\veloSum,\bm{y} |\veloIndexZero)}{p(\bm{y} |\veloIndexZero)}&=p(\veloSum|\bm{y} ,\veloIndexZero)
%\end{align}
When evaluating \eqref{chain_rule_cond_pdf} for the optimization of the conditional mutual information, the optimal decoder $p(\veloSum|\bm{y},\veloIndexZero)$ is intractable in practice \cite{simeone2022machine}. Thus, the mutual information cannot be maximized directly, but a lower bound on \eqref{chain_rule_cond_pdf} can be derived, which in turn can be optimized. 
One way of deriving such a lower bound is to use a variational approximation $q_{\bm{\phi}}(\veloSum|\bm{y},\veloIndexZero)$, with parameters ${\bm{\phi}}$, that approximates $p(\veloSum|\bm{y},\veloIndexZero)$ \cite{poole2019boundsMI,barber2004algorithm,edgar2023semantic}. 
With this approach, a lower bound on the conditional mutual information can be derived (given in Appendix \ref{Appendix:derivationMILowerbound}), which results in
\begin{align} \label{MILBo_short}
 &I(\veloSum;\bm{y} |\veloIndexZero) \geq
                             \\ &  - E_{p(\bm{y} ,\veloIndexZero)} \! \left[H  \left( p(\veloSum|\bm{y} ,\veloIndexZero), q_{\bm{\phi}}(\veloSum|\bm{y} ,\veloIndexZero)\right)\right] \nonumber  \\ & +   E_{p(\veloIndexZero)}  \left[H(p(\veloSum|\veloIndexZero)) \right]\nonumber .
\end{align}

%where $E_p[\cdot]$ is an expected value with respect to distribution $p$, $H(\cdot)$ is the Shannon entropy, $H(\cdot,\cdot)$ is the cross entropy, and $D_{\text{KL}}(p||q)$ is the Kullback-Leibler (KL) divergence from $p$ to $q$.
%The step for \eqref{MILBo} uses the fact that the KL divergence is non-negative \cite{cover1999informationTheory}. 

Instead of maximizing the conditional mutual information, now the lower bound in \eqref{MILBo_short} can be maximized. 
From the derivation it can be seen that the lower bound on the mutual information is tight, if the \ac{KL} divergence $D_{\text{KL}}\left( p(\veloSum|\bm{y} ,\veloIndexZero) || q_{\bm{\phi}}(\veloSum|\bm{y} ,\veloIndexZero) \right) = 0$, i.e., if the modeled probability distribution of the decoder $q_{\bm{\phi}}(\veloSum|\bm{y} ,\veloIndexZero)$ matches the true posterior distribution $p(\veloSum|\bm{y} ,\veloIndexZero)$ almost everywhere \cite{cover1999informationTheory}.
For an intuitive understanding of \eqref{MILBo_short} we can use 
\eqref{condMI_identity} from Appendix \ref{Appendix:derivationMILowerbound} to get
\begin{align}
&   -E_{p(\bm{y} ,\veloIndexZero)}  \left[H   \left( p(\veloSum|\bm{y} ,\veloIndexZero), q_{\bm{\phi}}(\veloSum|\bm{y} ,\veloIndexZero)\right)\right]    \nonumber \\ & +   E_{p(\veloIndexZero)}  \left[H(p(\veloSum|\veloIndexZero)) \right]\nonumber = \underbrace{I(\veloSum;\bm{y} |\veloIndexZero)}_\text{encoder objective} \\
    & - \underbrace{E_{p(\bm{y},\veloIndexZero)}  \left[D_{\text{KL}}\left( p(\veloSum|\bm{y} ,\veloIndexZero) || q_{\bm{\phi}}(\veloSum|\bm{y} ,\veloIndexZero) \right)\right]}_\text{decoder objective},
\end{align}
which shows that maximizing the conditional mutual information lower bound can be seen as optimizing an implicit encoder and decoder objective. The objective of the encoder is to maximize the mutual information of the encoded symbols after transmission over the channel $\bm{y}$ to be maximally informative about the semantic variable $\veloSum$ given the side information $\veloIndexZero$.
The objective of the decoder is to find the distribution $q_{\bm{\phi}}(\veloSum|\bm{y} ,\veloIndexZero)$ with the minimum \ac{KL} divergence to the true decoder distribution $p(\veloSum|\bm{y} ,\veloIndexZero)$.

For the optimization, the conditional entropy term $E_{p(\veloIndexZero)} \left[H(p(\veloSum|\veloIndexZero)) \right]$ can be omitted, as it is independent of the encoder and decoder parameters.
%Instead of maximizing the negative cross entropy, the cross entropy is minimized.
This gives us the following minimization problem of the encoder and decoder parameters ${\bm{\theta}_1},\dots,{\bm{\theta}_{\numberReceivers}}$ and ${\bm{\phi}}$, respectively, with
%of the encoders $p_{{\bm{\theta}_\idxNeigh}}(\bm{c}_\idxNeigh|\bm{\s}_\idxNeigh)$ and the decoder $q_{\bm{\phi}}(\veloSum|\bm{y})$ with 
%$\substack{p_{{\bm{\theta}_\idxNeigh}}(\bm{c}_\idxNeigh|\bm{\s}_\idxNeigh) \ \text{for}\ i=1,\dots,\numberReceivers  \\  q_{\bm{\phi}}(\veloSum|\bm{y}) \quad \quad \quad \quad\quad }$
\begin{align}  \label{MinCorssEntropy}
    \argmin_{\bm{\theta}_1,\dots,\bm{\theta}_{\numberReceivers},\bm{\phi}}  &E_{p(\bm{y} ,\veloIndexZero)}\left[H\left( p(\veloSum|\bm{y} ,\veloIndexZero), q_{\bm{\phi}}(\veloSum|\bm{y} ,\veloIndexZero)\right)\right] \\
    & \text{s.t.} \ \ \bm{y}  \in \mathbb{R}^{\numberReceivers\cdot N_{\mathrm{ch}}}, \, P_c \leq 1. \nonumber %\\ %\text{(limited bandwidth)} \nonumber \\
    %& \quad \quad  E[\bm{C}^2] \leq 1 \nonumber. %\,\text{dB} \ \text{(limited signal power)} .\nonumber 
\end{align}
%In the next section we discuss how to optimize the distribution $q_{\bm{\phi}}(\veloSum|\bm{y})$ with parameters ${\bm{\phi}}$ when optimizing \eqref{MinCorssEntropy} for the continuous distribution $p(\veloSum|\bm{y} ,\veloIndexZero)$.
Analogously, we can obtain the optimization problem to design the semantic communication system for the transmission of the velocity gradients. This gives us the following minimization problem of the encoder and decoder parameters ${\bm{\vartheta}_1},\dots,{\bm{\vartheta}_{\numberReceivers}}$ and ${\bm{\varphi}}$, respectively, with
\begin{align}  \label{MinCorssEntropyGradient}
    \argmin_{\bm{\vartheta}_1,\dots,\bm{\vartheta}_{\numberReceivers},\bm{\varphi}} \! &E_{p(\bm{y} ,\gradIndexZero)} \! \left[H\left( p(\gradSum|\bm{y} ,\gradIndexZero), q_{\bm{\varphi}}(\gradSum|\bm{y} ,\gradIndexZero)\right)\right] \\
    & \text{s.t.} \ \ \bm{y}  \in \mathbb{R}^{\numberReceivers\cdot N_{\mathrm{ch}}}, \, P_c \leq 1. \nonumber 
\end{align}
In the following section, we discuss how the minimization of \eqref{MinCorssEntropy} and \eqref{MinCorssEntropyGradient} can be implemented in practice using \acp{DNN}.

\subsection{Implementation Considerations}
%\subsection{Making the Optimization Problem Efficiently Solvable}
%In this section, it is discussed how the formulated optimization problem \eqref{MinCorssEntropy} can be  solved efficiently.

We want to optimize the parameters ${\bm{\theta}_1},\dots,{\bm{\theta}_{\numberReceivers}}$ and ${\bm{\phi}}$ of the encoders and the decoder for the transmission of the velocity models \eqref{MinCorssEntropy}, and
  the parameters $  {\bm{\vartheta}_1},\dots,{\bm{\vartheta}_{\numberReceivers}}$ and ${\bm{\varphi}}$ of the encoders and the decoder
for the transmission of the velocity gradients \eqref{MinCorssEntropyGradient}, respectively, of our \ac{DNN} communication systems. For this, training data is required, which we generate as described later in Section \ref{sec:training_data_generation}.
To optimize the \acp{DNN} parameters, we employ \ac{SGD} based optimization and leverage the reparameterization trick to enable efficient gradient estimation with respect to the encoder and decoder parameters 
(see Appendix~\ref{Implementation_considerations} for details).

%To train the \ac{NN}, training data samples of the observations $\veloIndexZero, \veloIndexOne,\dots,\veloIndexLast$ are used as the inputs, and training data samples of the semantic variable $\veloSum$ are used as the desired output. Furthermore, a suitable type of parameterized distribution for the decoder is selected such that \ac{NN} training is computationally feasible.

%In Appendix \ref{Implementation_considerations}, it is shown how the encoder and decoder parameters can be trained using the training data.
Furthermore, an efficiently computable loss function is required. However, \eqref{MinCorssEntropy} and \eqref{MinCorssEntropyGradient} are 
computationally infeasible due to high dimensionality of the distributions $q_{\bm{\phi}}(\veloSum|\bm{y} ,\veloIndexZero)$ and $q_{\bm{\varphi}}(\gradSum|\bm{y} ,\gradIndexZero)$ with parameters $\phi$ and $\varphi$, respectively.
In the following, we show that under the assumption of a Gaussian distribution for the decoder output distribution, the optimization problems of \eqref{MinCorssEntropy} and \eqref{MinCorssEntropyGradient} become computationally feasible. We show this in the following only for the case of the transmission of the velocity models, as it holds analogously for the case of the velocity gradients.
We emphasize here that this assumption is only required for the decoder distribution and is not required for the distribution of the input data.  

%In the following, a suitable type of parameterized distribution for the decoder is selected such that \ac{NN} training is computationally feasible.
%We note that this assumption of distribution is only required for the decoder output, not for the input data, i.e. we do not assume that the data to be transmitted follows 

In many works of semantic communication, where the task is to solve a classification problem, e.g., as in  \cite{edgar2023semantic,razlighi2024cooperative,razlighi2024semantic}, the posterior distribution of the decoder $p(\veloSum|\bm{y} ,\veloIndexZero)$ is a low dimensional discrete probability distribution. 
However, in our case, $p(\veloSum|\bm{y},\veloIndexZero)$ is a continuous, or more precisely, as we operate on float32 numbers on digital computer hardware, a high dimensional discrete distribution.
As an example, for a fairly moderate grid size of $100$ by $50$ float32 numbers for a subsurface image the number of events is about $2^{32 \cdot5000}$, which makes it impossible to estimate a probability for each event. Even assuming that the values at each grid point are independent, it is still infeasible, as there are approximately $5000\cdot2^{32}\approx 2\cdot10^{10}$ different discrete probability events.

As a consequence, simplification is required to estimate the probability density $q_{\bm{\phi}}(\veloSum|\bm{y},\veloIndexZero)$. Different parametric and non-parametric methods for probability density estimation methods exist \cite{silverman2018density}.
%Here, we use a common approach of simply assuming independent Gaussian distributions for each grid point with fixed variance. Our simulation results show that sufficiently good communication performance can be achieved under this assumption. We leave the investigation of different parametric distributions as further research work.
Here, we use a common approach of assuming independent Gaussian distributions for every $i$-th element $\veloSumi$ of $\veloSum\in\R^{\Nx\Nz}$ for $q_{\bm{\phi}}(\veloSum|\bm{y},\bm{\s}_{0})$ with fixed variance $\sigma^2$ and means $\bm{\mu}(\bm{y},\veloIndexZero)$. The $i$-th mean $\mu_{i}(\bm{y},\veloIndexZero)$ is the $i$-th output of the decoder $\bm{\mu}(\bm{y},\veloIndexZero)$, which depends on the decoder inputs $\bm{y}$ and $\veloIndexZero$. 
We mention that this assumption of the decoder distribution $q_{\bm{\phi}}(\veloSum|\bm{y},\bm{\s}_{0})$ essentially assumes the Gaussianity and uncorrelatedness of the residual errors of the estimate of the semantic variable $\veloSum$. In practice this might not be guaranteed to hold, however,
our simulation results show that good communication performance can be achieved under this assumption. We leave the investigation of different parametric distributions as further research work.

Given this assumption, our optimization objective of the cross entropy \eqref{MinCorssEntropy} becomes
% \begin{align}
%     & E_{p(\bm{y} ,\veloIndexZero)}\left[H\left( p(\veloSum|\bm{y} ,\veloIndexZero), q_{\bm{\phi}}(\veloSum|\bm{y} ,\veloIndexZero)\right)\right] \\
%     = &E_{p(\veloSum,\bm{y} ,\veloIndexZero)}\left[  \log\left( q_{\bm{\phi}}(\veloSum|\bm{y} ,\bm{\s}_{0})\right) \right]\\
%      = &E_{p(\veloSum,\bm{y} ,\veloIndexZero)}\left[  \log \!\left( \prod_{i=1}^{\Nx\Nz}  \frac{1}{\sqrt{2\pi\sigma^2}} \exp \! \left( -\frac{\left(z_i - \mu_{i}(\bm{y},\veloIndexZero)\right)^2}{2\sigma^2} \! \right) \! \right) \!\right]\\
%      = &E_{p(\veloSum,\bm{y} ,\veloIndexZero)} \!\!\left[ \! \frac{1}{2\sigma^2}\sum_{i=1}^{\Nx\Nz} \!\left(z_i - \mu_{i}(\bm{y},\veloIndexZero)\right)^2 \! \right] \!\! + \! \Nx\Nz\log \!\sqrt{2\pi\sigma^2} \\
%      =& \frac{1}{2\sigma^2} E_{p(\veloSum,\bm{y} ,\veloIndexZero)}\left[ \|\veloSum - \bm{\mu}(\bm{y},\veloIndexZero) \|_2^2 \right] + \Nx\Nz\log\sqrt{2\pi\sigma^2}.\label{objective_MSE}
% \end{align}
\begin{equation}
    \frac{1}{2\sigma^2} E_{p(\veloSum,\bm{y} ,\veloIndexZero)}\left[ \|\veloSum - \bm{\mu}(\bm{y},\veloIndexZero) \|_2^2 \right] + \Nx\Nz\log\sqrt{2\pi\sigma^2},\label{objective_MSE}
\end{equation}
with its derivation shown in Appendix \ref{Derivation_CE_under_Gaussian}. 
We  see that under the given assumption our optimization objective becomes the \ac{MSE} loss $E_{p(\veloSum,\bm{y} ,\veloIndexZero)}\left[ \|\veloSum - \bm{\mu}(\bm{y},\veloIndexZero) \|_2^2 \right]$ \cite{GoodfellowDeepLearning}.
%If the variance is fixed as a constant term, we see that  the loss function for the assumption of independent Gaussian distributions is the sample averaged \ac{MSE} loss .
Estimating the gradient for a batch of training data for \eqref{objective_MSE} with respect to the encoder and decoder parameters $\bm{\phi},\bm{\theta}_1,\dots,\bm{\theta}_{\numberReceivers}$ is done following \eqref{sample_average_loss_derivative} and \eqref{reparametrization_trick_applied} in Appendix~\ref{Appendix:nn_training}.
In an analogous way, the \ac{MSE} loss can be derived for the optimization problem \eqref{MinCorssEntropyGradient} to design the semantic communication system to exchange the velocity gradients. 
%An analogous derivation to obtain the MSE loss can be done for the cross entropy in \eqref{MinCorssEntropyGradient}.

%With these results, and the given assumptions, we can now maximize the conditional mutual information lower bound \eqref{MILBo_short} by applying the \ac{MSE} loss for the end-to-end training of the encoders and the decoder.
%In the following section, the encoders and decoder for different communication system designs are trained and evaluated with respect to the communication performance and 

\section{Performance Evaluation} 
\label{sec:simulation_results}
In the following, we evaluate the \ac{ATC-FWI} algorithm together with our proposed semantic \ac{DFC}.
First, we introduce the communication systems that serve as reference systems for our semantic \ac{DFC}. Then the channel models, the used \ac{NN} architectures, and the used training dataset are explained. In Section~\ref{simulation_results_communication} the communication performance is evaluated, and in Section~\ref{simulation_results_atc_fwi} the \ac{ATC-FWI} imaging performance is evaluated for the different communication systems.

\subsection{Description of Reference Systems}
Let us introduce the communication methods that serve as reference systems for our proposed semantic \ac{DFC}. The different systems are shown in Figs. \ref{fig:Digital} to \ref{fig:AirComp_Sem_D_JSCC}. To avoid redundancy, Fig. \ref{fig:JSCC_comparison} presents only the communication systems for exchanging the velocity models $\veloIndexEll$, as the setup for the velocity gradient exchange is analogous.
%We remind that two distinct communication steps happen in the \ac{ATC-FWI} algorithm, which are the transmissions of the  estimated velocity model$\vecModel$ and the gradients $\delta\vecModel$. For both transmission steps we use the proposed semantic communication approach.
We note again, that in the case of transmission of the velocity model, the observations are $\vecModel_\idxNeigh$, and the semantic variable is
$\veloSum = \sum_{\idxNeigh\in\setNeighNode} a_{\idxNeighNode} \widetilde{\vecModel}_\idxNeigh$.
For transmission of the gradients, the observations are $\delta\vecModel$, and the semantic variable is given as
$\gradSum = \sum_{\idxNeigh\in\setNeighNode} b_{\idxNeighNode} \delta\vecModel_\idxNeigh$.
In our simulations, we set the coefficients $a_{\idxNeighNode}$ and $b_{\idxNeighNode}$ to be $1/|\setNeighNode|$.
This means that the function $F$ in Figs. \ref{fig:Digital} to \ref{fig:D_JSCC} is defined to compute the arithmetic mean.

As simplest baseline scheme we employ a classical digital \ac{SSCC} transmission system and compare it to the proposed approaches, see Fig. \ref{fig:Digital}. For source coding of the classical communication system Huffman coding, PNG, or JPEG were considered, together with subsampling, quantization, and for the gradient dataset also cutting part of the image, to reduce the number of bits required to be sent over the channel. In our case with the size of the data $\Nx\Nz=5000$, PNG and JPEG were not suitable, as both require too much overhead data, making compression to very small number of channel uses not possible. We therefore use Huffmann coding for our evaluation.
For channel coding, we use a 5G LDPC code implementation from \cite{sionna}. For the modulation, $16$-QAM and $64$-QAM were considered. All parameters were optimized using grid search by using $10,000$ data samples from the training dataset to find the best parameter setting at the training SNR. The coding rate is selected such that the total number of channel uses of the digital and analog transmission of the \ac{JSCC} approaches is the same. 
In the distributed \ac{ATC-FWI} algorithm, data must be exchanged for every algorithm step, requiring immediate transmission. Therefore, each transmission has to be done in small blocks, which is known to limit the performance of \ac{SSCC} communication systems \cite{JSCC_survey}.

For a more suitable benchmark, a classic \ac{JSCC} architecture is used where no side information is used at the decoder, which is shown in Fig. \ref{fig:JSCC}. For the case of \ac{JSCC} without side information, each encoder/decoder pair is optimized to individually reconstruct the observation $\veloIndexIhat$ for the velocity model dataset or $\gradIndexIhat$ for the velocity gradient dataset.
To see how the availability of side information impacts the encoder/decoder design, we compare the case of \ac{JSCC} without side information to \ac{JSCC} with side information, which is shown in Fig. \ref{fig:JSCC_with_side_info}. All following methods use the side information available at the decoder.
For \ac{JSCC} (with side information), the encoder pairs are optimized to individually reconstruct the observation $\veloIndexIhat$ or $\gradIndexIhat$, respectively.
As we consider the side information for the encoder/decoder design for all further compared methods, we do not explicitly mention the consideration of side information to simplify the notation, and only mention it explicitly if no side information is used. 
The next approach, referred to as distributed \ac{JSCC} is shown in Fig. \ref{fig:D_JSCC}. It uses a joint decoder for all observations $\veloIndexOnehat, \dots, \veloIndexLasthat$ or $\gradIndexOnehat, \dots, \gradIndexLasthat$, respectively,
 %$\hat{\bm{\s}}_1, \dots, \hat{\bm{\s}}_{\numberReceivers}$
to exploit the correlation between the observations $1$ to $\numberReceivers$.

Furthermore, we compare the proposed semantic \ac{JSCC} and the semantic \ac{JSCC} \ac{OAC} methods, which are shown in Figs. \ref{fig:Sem_D_JSCC} and \ref{fig:AirComp_Sem_D_JSCC}, respectively. Both proposed methods directly reconstruct the semantic variables $\veloSum$ or $\gradSum$, respectively, as \ac{DFC}, instead of reconstructing the individual observations, as only the semantic variable is required for executing the \ac{ATC-FWI} algorithm. 
The benefit of the semantic \ac{JSCC} \ac{OAC} approach compared to the semantic \ac{JSCC} approach is that all agents can transmit on the same channel resources, potentially using the channel resources more efficiently. This comes with the constraint of formulating the transmitted data as a nomographic function, and with the drawback that synchronization is required \cite{OAC_survey1,OAC_survey2}.

\begin{figure}[t!]
\centering
\begin{subfigure}{0.49\textwidth}
    \centering
    \vspace{-0.15cm}
    \caption{Digital SSCC transmission (no side info)}
    \vspace{-0.12cm}
    \resizebox{\textwidth}{!}{
        \includegraphics{tikz/block_diagram/Digital.tikz}
    }
    \label{fig:Digital}
\end{subfigure}
\hfill
\vspace{-0.68cm}
\begin{subfigure}{0.49\textwidth}
    \centering
    \caption{JSCC (no side info)}
    \vspace{-0.12cm}
    \resizebox{\textwidth}{!}{
        \includegraphics{tikz/block_diagram/JSCC_y0.tikz}
    }
    \label{fig:JSCC}
\end{subfigure}
\hfill
\vspace{-0.68cm}
\begin{subfigure}{0.49\textwidth}
    \centering
    \caption{JSCC}
    \vspace{-0.12cm}
    \resizebox{\textwidth}{!}{
        \includegraphics{tikz/block_diagram/JSCC_side_information.tikz}
    }
    \label{fig:JSCC_with_side_info}
\end{subfigure}
\hfill
\vspace{-0.68cm}
\begin{subfigure}{0.49\textwidth}
    \centering
    \caption{Distributed JSCC}
    \vspace{-0.12cm}
    \resizebox{\textwidth}{!}{
        \includegraphics{tikz/block_diagram/D_JSCC_y0_backup.tikz}
    }
    \label{fig:D_JSCC}
\end{subfigure}
\hfill
\vspace{-0.68cm}
\begin{subfigure}{0.49\textwidth}
    \centering
    \caption{Semantic JSCC}
    \vspace{-0.12cm}
    \resizebox{\textwidth}{!}{
        \includegraphics{tikz/block_diagram/Sem_JSCC_y0_backup.tikz}
    }
    \label{fig:Sem_D_JSCC}
\end{subfigure}
\hfill
\vspace{-0.68cm}
\begin{subfigure}{0.49\textwidth}
    \centering
    \caption{Semantic JSCC \ac{OAC}}
    \vspace{-0.12cm}
    \resizebox{\textwidth}{!}{
        \includegraphics{tikz/block_diagram/OAC_JSCC_y0_backup.tikz}
    }
    \label{fig:AirComp_Sem_D_JSCC}
\end{subfigure}
\vspace{-0.6cm}
\caption{Comparison of transmission architectures for sharing neighboring velocity models $\veloIndexEll$ with agent $r$. Side information is always used, unless stated otherwise: (a) Digital transmission with \ac{SSCC} without side information (b) JSCC without side information, (c) JSCC, (d) distributed JSCC, (e) semantic JSCC, and (f) semantic JSCC \ac{OAC}.}
\label{fig:JSCC_comparison}

\end{figure}

\subsection{Channel Models} \label{sec:channel_models}
For the evaluation, we consider $\numberReceivers$  channels from each neighboring agents to the receiving agent as independent and identically distributed (i.i.d.) \ac{AWGN} channels. 
The average signal power per channel use at the decoder $P_c$ is given as $P_c = \frac{1}{\numberReceivers N_\text{ch}}\sum_{l=1}^{\numberReceivers N_\text{ch}} |c_i|^2 $, where $c_i$ is the $i$-th entry of the vector $\bm{c}$, which is given as
\begin{equation}
    \bm{c} = 
\begin{cases}
\bm{c}_1 +  \dots + \bm{c}_{\numberReceivers} \quad & \text{for \ac{OAC} channel} ,\\
\begin{bmatrix}
    \bm{c}_1^{\top} &  \dots & \bm{c}_{\numberReceivers}^{\top}
\end{bmatrix}^{\top}  & \text{otherwise} .
\end{cases}
\end{equation}
For the \ac{OAC} channel, the power of the additive noise is $\sigma_n^2$. 
Otherwise, for the $\numberReceivers$ channels we assume noise powers of $\sigma_{n_1}^2 = \dots = \sigma^2_{n_{\numberReceivers}} = \sigma_n^2$.
The noise power $\sigma_n^2$ is scaled according to
$\text{SNR} = \frac{P_c}{\sigma^2_n}$.

Finally, the total number of channel uses for neighboring agents transmitting to the receiving agent is always $\numberReceivers N_\text{ch}$, i.e., for the case \ac{OAC} channel, each user has $\numberReceivers N_\text{ch}$ channel uses, and otherwise each of the $\numberReceivers$ transmitting agents has $N_\text{ch}$ channel uses.
For \ac{OAC},  we neglect synchronization offset errors in this work, thus providing an upper bound on the communication performance of \ac{OAC}.
We mention that many works exist on agent synchronization from the research area of collaborative distributed beamforming, where precise synchronization is required as in \ac{OAC} \cite{jayaprakasam2017distributed}.
Synchronization among agents is commonly achieved using iterative algorithms that incorporate feedback from the receiver \cite{jayaprakasam2017distributed}.
However, we still note that in future work, the effect of synchronization errors should be investigated for semantic communication with \ac{OAC}.

\subsection{Neural Network Architecture}
For the encoder and decoder we use a \ac{CNN} architecture \cite{li2021survey}. Each encoder consists of five 2D-convolution layers with $60$ $5\times5$ convolution filters each with batch normalization and PReLU activation function, with respective strides of $\{2,2,2,1,1\}$, followed by $3$ fully connected layers with PReLU activations except for the last layer, where no activation function is used. The same setup is used for the decoder, but at the start $3$ fully connected layers are used, followed by transposed convolution layers with respective strides of $\{1,1,2,2,2\}$. The number of filters of the last encoder layer is selected to match the number of channel uses $N_\text{ch}$, and the last decoder layer has just one filter to output a $100$ by $50$ grid of the velocity model/the gradients. Furthermore, data normalization is done before the first convolution layer and denormalization after the last decoder layer. To match the average power constraint power normalization is done after the last encoder layer.

For simplification, we use parameter sharing for the encoders and decoders, such that not all combinations of sending and receiving agents need to be trained individually.
For the transmissions of the velocity models and the gradients the same neural network architecture is used, but the training is done with different datasets.

\subsection{Training Data Generation and Training} 
\label{sec:training_data_generation}
As we use \acp{NN} for the encoder and decoder and do end-to-end training, we require training data. To generate the training data, we first generate several scenarios of velocity models as ground truth $P$-wave velocity models.
We consider a $\SI{200}{\meter}$ wide and $\SI{100}{\meter}$ deep two dimensional slice of the subsurface with associated velocity values at grid points every two meters in both directions.
We consider three different scenarios of ground truth velocity models. The first two scenarios are rectangular and elliptically shaped velocity anomalies with a constant background, and the third scenario consists of three layers with varying width with different velocities.

To generate the training data for the communication system consisting of the velocity models $\widetilde{\vecModel}_\idxNeigh$ and the velocity gradients $\delta\vecModel_\idxNeigh$ for all agents $\idxNeigh$, the \ac{ATC-FWI} algorithm is executed with error-free communication links, and the respective velocity models and gradients are stored as training data.

$14 \, 000$ different velocity ground truth models were generated, and the \ac{ATC-FWI} algorithm was run for $20$ iterations with an initial model of a Gaussian filtered ground truth for each scenario. For the training in total only $80\ 000$ training data samples were used due to memory constraints.
For the velocity dataset, samples were removed where all agents had nearly identical data to avoid overfitting to the side information, and for the gradient dataset $80\ 000$ samples were selected randomly.
All models are trained for $300$ epochs with a learning rate of  $10^{-4}$. For the last few epochs the learning rate is lowered stepwise until $10^{-7}$.

% Then the \ac{ATC-FWI} algorithm is run with perfect communication to generate the velocity dataset consisting of the velocity estimates $\widetilde{\vecModel}_\idxNeigh$ and the gradient dataset consisting of the gradients $\delta\vecModel_\idxNeigh$.
% The ground truth of the semantic variable for each instance is calculated for the velocity dataset with $\sum_{\idxNeigh\in\setNeighNode} a_{\idxNeighNode} \widetilde{\vecModel}_\idxNeigh$, where we set $a_{\idxNeighNode} = \frac{1}{|\setNeighNode|}$.

\subsection{Results - Communication Performance} \label{simulation_results_communication}

\renewcommand{\xminimum}{0}
\renewcommand{\xmaximum}{20}
\renewcommand{\yminimum}{-70}
\renewcommand{\ymaximum}{-10}
\renewcommand{\xticklabelList}{0,5,10,15, 20}
\renewcommand{\yticklabelList}{-70, -65, -60,-55,-50,-45,-40,-35,-30,-25,-20,-15,-10,-5,0}
\renewcommand{\xlabel}{SNR [dB]}
\renewcommand{\ylabel}{NMSE [dB]}

\newcommand{\customfontsize}{\fontsize{19}{24}\selectfont}
\newcommand{\nodedistance}{2.5cm}

\renewcommand{\legendfontsize}{\normalsize}
\renewcommand{\ltickfontsize}{\normalsize} % normalsize
\renewcommand*{\labelaxisfontsize}{\large}

\setlength{\figH}{6.35cm} %6.35
\setlength{\figW}{4.85cm} %4.85cm
\renewcommand*{\linewidthtikz}{2.2pt}
\renewcommand*{\marksize}{3.0} %2.6

\begin{figure*}[t!]
\centering

% ================= Legend =================
\vspace{-1.5cm}
\hspace{-20em}
\resizebox{\textwidth}{!}{
    \begin{tikzpicture}
\begin{axis}[ % 
 hide axis, 
xmin = 10, 
xmax = 50, 
ymin = 0, 
ymax = 0.4, 
legend columns = 2, \legendtranspose
legend style={at={(1.03,0.5)},font=\legendfontsize, 
anchor = south west, 
draw = white!15!black, 
legend cell align = left}] 

\addlegendimage{color={rgb, 1: red, 0.0;green, 0.4666666666666667;blue, 0.7333333333333333}, mark size=\marksize, mark=triangle, mark options={ solid,rotate=180, color={rgb, 1: red, 0.0;green, 0.4666666666666667;blue, 0.7333333333333333}},, line width=\linewidthtikz}
\addlegendentry{\ifthenelse{\boolean{LegendAutomatic}}{Digital (no side info)}{\legendlabelf} }
 
\addlegendimage{color={rgb, 1: red, 0.8666666666666667;green, 0.8;blue, 0.4666666666666667}, mark size=\marksize, mark=o, mark options={ solid,color={rgb, 1: red, 0.8666666666666667;green, 0.8;blue, 0.4666666666666667}},, line width=\linewidthtikz}
\addlegendentry{\ifthenelse{\boolean{LegendAutomatic}}{JSCC (no side info)}{\legendlabela} }

\addlegendimage{color={rgb, 1: red, 0.9333333333333333;green, 0.4666666666666667;blue, 0.2}, mark size=\marksize, mark=diamond, mark options={ solid,color={rgb, 1: red, 0.9333333333333333;green, 0.4666666666666667;blue, 0.2}},, line width=\linewidthtikz}
\addlegendentry{\ifthenelse{\boolean{LegendAutomatic}}{JSCC}{\legendlabelb} }
 
\addlegendimage{color={rgb, 1: red, 0.06666666666666667;green, 0.4666666666666667;blue, 0.2}, mark size=\marksize, mark=triangle, mark options={ solid,color={rgb, 1: red, 0.06666666666666667;green, 0.4666666666666667;blue, 0.2}},, line width=\linewidthtikz}
\addlegendentry{\ifthenelse{\boolean{LegendAutomatic}}{Distributed JSCC}{\legendlabelb} }
 
\addlegendimage{color={rgb, 1: red, 0.6666666666666666;green, 0.26666666666666666;blue, 0.6}, mark size=\marksize, mark=square, mark options={ solid,color={rgb, 1: red, 0.6666666666666666;green, 0.26666666666666666;blue, 0.6}},, line width=\linewidthtikz}
\addlegendentry{\ifthenelse{\boolean{LegendAutomatic}}{Semantic JSCC}{\legendlabelc} }
 
\addlegendimage{color={rgb, 1: red, 0.2;green, 0.13333333333333333;blue, 0.5333333333333333}, mark size=\marksize, mark=asterisk, mark options={ solid,color={rgb, 1: red, 0.2;green, 0.13333333333333333;blue, 0.5333333333333333}},, line width=\linewidthtikz}
\addlegendentry{\ifthenelse{\boolean{LegendAutomatic}}{Semantic JSCC AirComp}{\legendlabeld} }
 
\end{axis} 
\end{tikzpicture}
}

\vspace{-8.5em}

\vspace{-0.3cm}

\captionsetup[subfigure]{margin={2em,0pt}}
% ================= Row 1 =================
\begin{subfigure}{0.24\textwidth}
    \caption{Velocity models dataset \\ $N_\text{ch}=91$, $\numberReceivers=2$}
    \includegraphics[width=1\linewidth]{tikz/Comm_performance/over_SNR_velocity.tikz}
    \label{fig:comm_vel_91}
\end{subfigure}
\hfill
\begin{subfigure}{0.24\textwidth}
    \centering
    \caption{Velocity models dataset \\ $N_\text{ch}=45$, $\numberReceivers=2$}
    \includegraphics[width=1\linewidth]{tikz/Comm_performance/45_over_SNR_velocity.tikz}
    \label{fig:comm_vel_45}
\end{subfigure}
\hfill
\begin{subfigure}{0.24\textwidth}
    \centering
    \caption{Velocity models dataset \\ $N_\text{ch}=91$, $\numberReceivers=7$}
    \includegraphics[width=1\linewidth]{tikz/Comm_performance/over_SNR_fc_velocity.tikz}
    \label{fig:comm_vel_91fc}
\end{subfigure}
\hfill
\begin{subfigure}{0.24\textwidth}
    \centering
    \caption{Velocity models dataset \\ $N_\text{ch}=45$, $\numberReceivers=7$}
    \includegraphics[width=1\linewidth]{tikz/Comm_performance/45_over_SNR_fc_velocity.tikz}
    \label{fig:comm_vel_45fc}
\end{subfigure}

\vspace{0.6em}

% ================= Row 2 =================
\renewcommand{\xminimum}{0}
\renewcommand{\xmaximum}{20}
\renewcommand{\yminimum}{-22}
\renewcommand{\ymaximum}{-0}
\renewcommand{\xticklabelList}{-20,-10,0,5,10,15, 20}
\renewcommand{\yticklabelList}{-35,-30,-25,-22,-20,-18,-16,-14,-12, -10,-8,-6,-4,-2,0,10,20, 30}

\vspace{-0.5cm}

\begin{subfigure}{0.245\textwidth}
    \centering
    \caption{Velocity gradients dataset \\ $N_\text{ch}=91$, $\numberReceivers=2$}
    \includegraphics[width=1\linewidth]{tikz/Comm_performance/over_SNR_gradient.tikz}
    \label{fig:comm_grad_91}
\end{subfigure}
\hfill
\begin{subfigure}{0.245\textwidth}
    \centering
     \caption{Velocity gradients dataset \\ $N_\text{ch}=45$, $\numberReceivers=2$}
    \includegraphics[width=1\linewidth]{tikz/Comm_performance/45_over_SNR_gradient.tikz}
    \label{fig:comm_grad_45}
\end{subfigure}
\hfill
\begin{subfigure}{0.245\textwidth}
    \centering
    \caption{Velocity gradients dataset \\ $N_\text{ch}=91$, $\numberReceivers=7$}
    \includegraphics[width=1\linewidth]{tikz/Comm_performance/over_SNR_fc_gradient.tikz}
    \label{fig:comm_grad_91fc}
\end{subfigure}
\hfill
\begin{subfigure}{0.245\textwidth}
    \centering
    \caption{Velocity gradients dataset \\ $N_\text{ch}=45$, $\numberReceivers=7$}
    \includegraphics[width=1\linewidth]{tikz/Comm_performance/45_over_SNR_fc_gradient.tikz}
    \label{fig:comm_grad_45fc}
\end{subfigure}

\vspace{-0.3cm}

\caption{Communication performance: NMSE of the semantic variable in dB for different channel SNRs is shown. The training SNR is $10\,\text{dB}$. 
The results are shown for the velocity model evaluation dataset (top half) and the velocity gradient evaluation dataset (bottom half). The number of senders is $\numberReceivers=2$ on the left and $\numberReceivers=7$ on the right.
The number of channel uses is $N_\text{ch}=91$ for (a), (c), (e), and (g), and $N_\text{ch}=45$ otherwise.}
\label{fig:all_plots}
\end{figure*}

\renewcommand*{\linewidthtikz}{1.0pt}
\renewcommand*{\marksize}{2.0} %2.6
\renewcommand*{\marksizeRed}{0.8}
\renewcommand*{\marksizeBlue}{2.0}

\setlength{\figH}{2.85cm} %2.85cm
\setlength{\figW}{4.85cm} %4.85cm

To evaluate the communication performance for the velocity model dataset, we employ the \ac{NMSE} of the semantic variable 
$\frac{\| \veloSumEstimate - \veloSum\|_2^2}{\| {\veloSum}\|_2^2}$ where  $\veloSumEstimate$ is the reconstructed semantic variable at the receiving agent, and $\veloSum$ is the ground truth semantic variable. Analogously, to evaluate the communication performance for the velocity gradient dataset, we employ the \ac{NMSE} of the semantic variable 
$\frac{\| \gradSumEstimate - \gradSum\|_2^2}{\| {\gradSum}\|_2^2}$, where  $\gradSumEstimate$ is the reconstructed semantic variable at the receiving agent, and $\gradSum$ is the ground truth semantic variable.
Fig. \ref{fig:all_plots} shows the \ac{NMSE} of the semantic variable for all combinations of $\numberReceivers=2$, $\numberReceivers=7$ and $N_\text{ch}=91$, $N_\text{ch}=45$ over different channel SNRs for $15\,000$ evaluation data samples of the velocity model and velocity gradient datasets that were not used for training.
The training SNR is set to $10\,\text{dB}$.

%Fig. \ref{fig:velocity_line_overSNR} shows the results for the velocity dataset, and Fig. \ref{fig:gradient_line_over_SNR} shows the results for the gradient dataset for the case of two transmitting agents. Analogously, Fig. \ref{fig:fc_over_SNR} shows the results for seven transmitting agents.

For the velocity dataset, the results are shown in Fig.~\ref{fig:all_plots}\subref{fig:comm_vel_91}--\subref{fig:comm_vel_45fc}. It is observed that considering the side information at the decoder significantly improves the \ac{NMSE} of the semantic variable for the velocity dataset, especially for $\numberReceivers=7$. This is expected as the correlation between the velocity models from the transmitting agents to the receiving agent is high. 
The distributed \ac{JSCC}, the semantic \ac{JSCC}, and the semantic \ac{JSCC} \ac{OAC} approach achieve a lower \ac{NMSE} than the \ac{JSCC} approach, as they consider the correlation of the velocity models between all agents. The three models all achieve nearly the same \ac{NMSE}.

The achieved \ac{NMSE} for the semantic \ac{JSCC} approach can be expected to be the same as the distributed \ac{JSCC} approach, as the pixel-wise correlation of the velocity models is always non-negative. 
%An explanation why the achieved \ac{NMSE} for the semantic \ac{JSCC}  approach is nearly the same than the distributed \ac{JSCC} approach can be found for the specific case of Gaussian sources for the distributed linear function computation problem.
At least for the case of linear function computation for Gaussian sources, the following result is available in the literature:
When the linear function is the sum of two variables and the correlation between the two variables is non-negative, the required sum rate for asymptotically error-free reconstruction is the same regardless of whether the individual variables are reconstructed or the sum is directly reconstructed \cite{Gaussian_Two_Encoder_Source_Coding_Comparison, Distributed_compression_gap_to_Optimal_Sum_Rate}.
Furthermore, if the correlation between the two variables is negative, the required sum rate is smaller if the decoder directly reconstructs the function \cite{Distributed_compression_gap_to_Optimal_Sum_Rate}.

% At least for Gaussian sources it is known that for linear function computation of two variables, which can be written as a sum of two variables with positive weights, the required sum rate for asymptotically error free reconstruction is the same whether the linear function is reconstructed or the individual variables are reconstructed, if the correlation between the two variables is non-negative \cite{Gaussian_Two_Encoder_Source_Coding_Comparison, Distributed_compression_gap_to_Optimal_Sum_Rate} \todo{sentence too long, can you reformulate? Not easy to follow, maybe use two sentences}. 
% Furthermore, if the correlation between the two variables is negative, the required sum rate is smaller if the decoder directly reconstructs the function \cite{Distributed_compression_gap_to_Optimal_Sum_Rate}.

%--------------Plot ATC-FWI  ------------ ----------------------
\renewcommand{\xminimum}{0}
\renewcommand{\xmaximum}{40}
\renewcommand{\ymaximum}{0.042}
\renewcommand{\yminimum}{0.0315}
\renewcommand{\xlabel}{Iteration}
\renewcommand{\ylabel}{NMSE}
\renewcommand{\xticklabelList}{0, 10,20,30,40}
\renewcommand{\yticklabelList}{0.042,0.040,0.038,0.036,0.034,0.032, 0.030, 0.028, 0.026,0.024}

\renewcommand{\markrepeat}{3}

\renewcommand{\legendfontsize}{\small}
\renewcommand{\ltickfontsize}{\normalsize} % normalsize
\renewcommand*{\labelaxisfontsize}{\large}

\setlength{\figH}{6.35cm} %2.85cm
\setlength{\figW}{4.85cm} %4.85cm
\renewcommand*{\linewidthtikz}{2.2pt}
\renewcommand*{\marksize}{3.0} %2.6

\begin{figure}[t]
    \centering
    %\begin{subfigure}{1.0\columnwidth}
		%\includegraphics[width=1.0\linewidth]{tikz/atc_fwi_eval/atc_fwi_legend.tikz}
    %\end{subfigure}
    \vspace{-3.5cm}
    \begin{tikzpicture}
\begin{axis}[ % 
 hide axis, 
xmin = 10, 
xmax = 50, 
ymin = 0, 
ymax = 0.4, 
legend columns = 2, \legendtranspose
legend style={at={(0.0,0.05)},font=\legendfontsize, 
anchor = south west, 
draw = white!15!black, 
legend cell align = left}]

\addlegendimage{color={rgb, 1: red, 0.0;green, 0.4666666666666667;blue, 0.7333333333333333}, mark size=\marksize, mark=triangle, mark options={ solid,rotate=180, color={rgb, 1: red, 0.0;green, 0.4666666666666667;blue, 0.7333333333333333}},, line width=\linewidthtikz}
\addlegendentry{\ifthenelse{\boolean{LegendAutomatic}}{Digital (no side info)}{\legendlabelf} }

\addlegendimage{color={rgb, 1: red, 0.8666666666666667;green, 0.8;blue, 0.4666666666666667}, mark size=\marksize, mark=o, mark options={ solid,color={rgb, 1: red, 0.8666666666666667;green, 0.8;blue, 0.4666666666666667}},, line width=\linewidthtikz}
\addlegendentry{\ifthenelse{\boolean{LegendAutomatic}}{JSCC (no side info)}{\legendlabela} }

\addlegendimage{color={rgb, 1: red, 0.9333333333333333;green, 0.4666666666666667;blue, 0.2}, mark size=\marksize, mark=diamond, mark options={ solid,color={rgb, 1: red, 0.9333333333333333;green, 0.4666666666666667;blue, 0.2}},, line width=\linewidthtikz}
\addlegendentry{\ifthenelse{\boolean{LegendAutomatic}}{JSCC}{\legendlabelb} }
 
\addlegendimage{color={rgb, 1: red, 0.06666666666666667;green, 0.4666666666666667;blue, 0.2}, mark size=\marksize, mark=triangle, mark options={ solid,color={rgb, 1: red, 0.06666666666666667;green, 0.4666666666666667;blue, 0.2}},, line width=\linewidthtikz}
\addlegendentry{\ifthenelse{\boolean{LegendAutomatic}}{Distributed JSCC}{\legendlabelb} }
 
\addlegendimage{color={rgb, 1: red, 0.6666666666666666;green, 0.26666666666666666;blue, 0.6}, mark size=\marksize, mark=square, mark options={ solid,color={rgb, 1: red, 0.6666666666666666;green, 0.26666666666666666;blue, 0.6}},, line width=\linewidthtikz}
\addlegendentry{\ifthenelse{\boolean{LegendAutomatic}}{Semantic JSCC}{\legendlabelc} }
 
\addlegendimage{color={rgb, 1: red, 0.2;green, 0.13333333333333333;blue, 0.5333333333333333}, mark size=\marksize, mark=asterisk, mark options={ solid,color={rgb, 1: red, 0.2;green, 0.13333333333333333;blue, 0.5333333333333333}},, line width=\linewidthtikz}
\addlegendentry{\ifthenelse{\boolean{LegendAutomatic}}{Semantic JSCC AirComp}{\legendlabeld} }
 
\addlegendimage{color={rgb, 1: red, 0.5333333333333333;green, 0.13333333333333333;blue, 0.3333333333333333}, mark size=\marksize, mark=Mercedes star flipped, mark options={ solid,color={rgb, 1: red, 0.5333333333333333;green, 0.13333333333333333;blue, 0.3333333333333333}},, line width=\linewidthtikz}
\addlegendentry{\ifthenelse{\boolean{LegendAutomatic}}{Error-free communication}{\legendlabele} }
 
\end{axis} 
\end{tikzpicture}
    \vspace{-0.3cm}
    
    \begin{subfigure}{0.49\columnwidth}
        \centering
        \caption{Two comm. neighbors}
        \includegraphics[width=1.0\columnwidth]{tikz/atc_fwi_eval/atc_fwi_over_iteration91.tikz}
        \label{fig:nmse_over_iteration}
    \end{subfigure}
    \hfill
    \begin{subfigure}{0.49\columnwidth}
        \centering
        \caption{Full mesh comm. topology}
        \renewcommand{\ymaximum}{0.042}
        \renewcommand{\yminimum}{0.025}
        \includegraphics[width=1.0\columnwidth]{tikz/atc_fwi_eval/atc_fwi_over_iteration_fc91.tikz}
        \label{fig:nmse_over_iteration_fc}
    \end{subfigure}

    \vspace{-0.3cm}

    \begin{subfigure}{0.49\columnwidth}
        \centering
        \caption{Two comm. neighbors}
        \renewcommand{\yminimum}{0}
            \renewcommand{\ymaximum}{5}
            \renewcommand{\xlabel}{Iteration}
            \renewcommand{\ylabel}{Cost $\mathcal{L} (\model)$ in \eqref{eq:global_cost}}
            \renewcommand{\yticklabelList}{0, 1, 2, 3, 4,5}
			\includegraphics[width=1.0\columnwidth]{tikz/atc_fwi_eval/atc_fwi_over_iteration_cost91.tikz}
            \label{fig:cost_over_iteration}
    \end{subfigure}
    \hfill
    \begin{subfigure}{0.49\columnwidth}
        \centering
        \caption{Full mesh comm. topology}
        \renewcommand{\yminimum}{0}
            \renewcommand{\ymaximum}{5}
            \renewcommand{\xlabel}{Iteration}
            \renewcommand{\ylabel}{Cost $\mathcal{L} (\model)$ in \eqref{eq:global_cost}}
            \renewcommand{\yticklabelList}{0, 1, 2, 3, 4,5}
			\includegraphics[width=1.0\columnwidth]{tikz/atc_fwi_eval/atc_fwi_over_iteration_cost_fc91.tikz}
            \label{fig:cost_over_iteration_fc}
    \end{subfigure}

    \vspace{-0.3cm}
    
    \caption{\ac{ATC-FWI} algorithm performance: 
    NMSE of the estimated velocity averaged over all agents and the \ac{ATC-FWI} cost function over the algorithm iterations is shown for a MAS of eight agents with a line communication topology, i.e. two communicating neighbors for each agent, except at the edges only one communicating neighbor (a), (c), and with a full mesh communication topology, i.e. seven communicating neighbors for each agent (b), (d).}
    \label{fig:four-subplots}
\end{figure}

For the gradient dataset, where the results are shown in Fig.~\ref{fig:all_plots}\subref{fig:comm_grad_91}--\subref{fig:comm_grad_45fc}, the absolute value of the correlation of the gradients to be transmitted is lower compared to the velocity dataset. However, there is some negative pixel-wise correlation for the gradient dataset. 
The digital transmission optimized for $10\,\text{dB}$ SNR fails completely for smaller SNRs. The required compression from $\Nx\Nz=5000$ grid point values with $32$ bits float values per pixel down to $N_\text{ch}=91$ or $N_\text{ch}=45$ channel uses is not possible with high accuracy.
For the gradient dataset, using the side information at the decoder resulted in overfitting of the encoders and decoder during training. As a consequence, the side information was not used for the \ac{JSCC} approach. For the gradient dataset, the improvement in \ac{NMSE} of the semantic \ac{JSCC} approach to the distributed \ac{JSCC} approach is significantly larger compared to the case of the velocity dataset. As discussed above, for the special case of Gaussian sources, the required sum rate can be smaller if the decoder directly reconstructs the function in the semantic \ac{JSCC} approach compared to the distributed \ac{JSCC} approach, where all individual variables are reconstructed if the correlation is negative. This effect is observed to be larger for the case of seven transmitters in Fig.~\ref{fig:all_plots}\subref{fig:comm_grad_91fc}--\subref{fig:comm_grad_45fc}.
Finally, the semantic \ac{JSCC} \ac{OAC} approach outperforms all other methods for the gradient dataset, which is expected here, as \ac{OAC} is more advantageous for a larger number of agents, as the shared bandwidth for all agents increases \cite{abari2016over}.

\subsection{Results - \ac{ATC-FWI} Algorithm Performance With Communication} \label{simulation_results_atc_fwi}
We now integrate the various communication systems from the previous section into the \ac{ATC-FWI} and evaluate its overall imaging performance.
% For the simulation of the \ac{ATC-FWI} algorithm with the different communication systems, we use the trained end-to-end communication systems that were evaluated for their communication performance in the previous section. 
Furthermore, we compare our proposed semantic \ac{DFC} systems with the case where no errors during communication occur, which is referred to as error-free communication in the following. 

Our evaluation metric for the \ac{ATC-FWI} algorithm at iteration  $\iter$ is the NMSE of the estimated velocity averaged over all $\noRec$ agents, given as
\begin{equation}
    \frac{1}{\noRec}\sum_{\idxRec=1}^\noRec   \frac{ \| \vecModel_\idxRec^{[\iter]}- \vecModel^*\|_2^2 }{\| \vecModel^* \|_2^2},
\end{equation}
where $\vecModel^*$ is the velocity model ground truth. Furthermore, we show the global cost \eqref{eq:global_cost} of the \ac{ATC-FWI} algorithm over the number of iterations $\iter$. The results are averaged over $100$ independent runs with different velocity ground truths from the evaluation dataset.
The seismic source locations are fixed with equidistant positions for all simulations with $\noSrc=16$. We employ $\noRec=8$ agents for all simulations, where for each run, the agents' positions are selected randomly on a line of the surface of the two dimensional slice of the ground. We investigate a line and a full mesh communication topology. 
For the line topology every agent is able to communicate only with its direct neighbors, i.e, each agent can communicate with a maximum of two neighbors. This corresponds to the investigation in the previous section with two transmitting agents. Note that the agents at the end of the line array can only communicate with one neighbor.
For the full mesh topology, every agent can communicate with every other agent. With eight agents, this corresponds to the investigation in the previous section with seven transmitting neighbors.

The \ac{ATC-FWI} algorithm requires an adequate initial velocity model. To this end, we use a Gaussian smoothed version of the true velocity model. In practice, distributed traveltime tomography algorithm from \cite{Shin2022} can be used to obtain a low resolution subsurface reconstruction that serves as an initial model.
Furthermore, we locally stop the \ac{ATC-FWI} algorithm at each agent if the local cost \eqref{eq:local_cost} increases by any value larger than 0 compared to the local cost of the previous iteration, to avoid divergence of the algorithm. 
Such behavior can arise from errors occurring in the communication between agents.
Figs. \ref{fig:cost_over_iteration} and \ref{fig:cost_over_iteration_fc} depict the cost of the \ac{ATC-FWI} over the iterations.

\subsubsection{Results over number of iterations}
The results for the line topology, where the agents can only communicate with their direct neighbors for $N_\text{ch}=91$, are shown in Fig. \ref{fig:nmse_over_iteration}. The \ac{SSCC} digital communication scheme fails completely for the compression from $\Nx\Nz=5000$ grid point values with $32$ bits per pixel to $N_\text{ch}=91$ channel uses. The transmission errors are larger than the \ac{NMSE} at initialization leading to algorithm divergence.
If no side information is used at the decoder, the performance is also significantly limited by the compression to $N_\text{ch}=91$ channel uses. 
The semantic \ac{JSCC} and the semantic \ac{JSCC} \ac{OAC} perform similarly and slightly outperform the other methods in terms of NMSE of the estimated velocity model.

For a full mesh topology (Fig. \ref{fig:nmse_over_iteration_fc}) with $N_\text{ch}=91$, the overall \ac{NMSE} is significantly lower compared to the line topology after $40$ iterations, as the gradients and velocity models can be exchanged more easily between the agents. Furthermore, the advantage of the semantic \ac{JSCC} and the semantic \ac{JSCC} \ac{OAC} approaches is larger and they approach the performance of the case with error-free communication, which can be explained by the previously observed lower \ac{NMSE} for the gradient transmission in Fig. \ref{fig:all_plots}. 

\subsubsection{Results over SNR}

We depict the results of the ATC-FWI algorithm performance for different channel SNRs using a full mesh topology in Fig. \ref{fig:atc_fwi_over_SNR}. The results show the algorithm behavior after $35$ iterations. 
Again, the results are averaged over $100$ independent runs with different velocity ground truths from the evaluation dataset.
The two cases of $N_\text{ch}=91$ (Fig. \ref{fig:over_snr_91fc}) and $N_\text{ch}=45$ (Fig. \ref{fig:over_snr_45fc}) are compared to each other. For a larger SNR the NMSE is smaller, which is as expected. Furthermore, comparing the cases of $N_\text{ch}=91$ to $N_\text{ch}=45$, it can be seen that the advantage of the semantic \ac{JSCC} and semantic \ac{JSCC} \ac{OAC} is larger compared to the other methods, especially to distributed \ac{JSCC}, for $N_\text{ch}=45$, which indicates that the semantic approach works well for very high compression.

%--------------Plot ATC-FWI Over SNR ------------ ----------------------
\renewcommand{\xminimum}{0}
\renewcommand{\xmaximum}{15}
\renewcommand{\ymaximum}{0.042}
\renewcommand{\yminimum}{0.025}
\renewcommand{\xlabel}{SNR [dB]}
\renewcommand{\ylabel}{NMSE}
\renewcommand{\xticklabelList}{0,5, 10,15,20,30,40}
\renewcommand{\yticklabelList}{0.042,0.040,0.038,0.036,0.034,0.032, 0.030,0.028, 0.026,0.024}

\renewcommand{\markrepeat}{1}

\renewcommand{\legendfontsize}{\small}
\renewcommand{\ltickfontsize}{\normalsize} % normalsize
\renewcommand*{\labelaxisfontsize}{\large}

\setlength{\figH}{6.35cm} %2.85cm
\setlength{\figW}{4.85cm} %4.85cm
\renewcommand*{\linewidthtikz}{2.2pt}
\renewcommand*{\marksize}{3.0} %2.6

\begin{figure}[t]
    \centering
  %   \begin{subfigure}{0.9\columnwidth}
		% \includegraphics[width=0.9\linewidth]{tikz/atc_fwi_eval/atc_fwi_legend_short.tikz}
  %   \end{subfigure}

    \vspace{-3.9cm}
    \begin{tikzpicture}
\begin{axis}[ % 
 hide axis, 
xmin = 10, 
xmax = 50, 
ymin = 0, 
ymax = 0.4, 
legend columns = 2, \legendtranspose
legend style={at={(0.00,0.05)},font=\legendfontsize, 
anchor = south west, 
draw = white!15!black, 
legend cell align = left}] 

\addlegendimage{color={rgb, 1: red, 0.8666666666666667;green, 0.8;blue, 0.4666666666666667}, mark size=\marksize, mark=o, mark options={ solid,color={rgb, 1: red, 0.8666666666666667;green, 0.8;blue, 0.4666666666666667}},, line width=\linewidthtikz}
\addlegendentry{\ifthenelse{\boolean{LegendAutomatic}}{JSCC (no side info)}{\legendlabela} }

\addlegendimage{color={rgb, 1: red, 0.9333333333333333;green, 0.4666666666666667;blue, 0.2}, mark size=\marksize, mark=diamond, mark options={ solid,color={rgb, 1: red, 0.9333333333333333;green, 0.4666666666666667;blue, 0.2}},, line width=\linewidthtikz}
\addlegendentry{\ifthenelse{\boolean{LegendAutomatic}}{JSCC}{\legendlabelb} }
 
\addlegendimage{color={rgb, 1: red, 0.06666666666666667;green, 0.4666666666666667;blue, 0.2}, mark size=\marksize, mark=triangle, mark options={ solid,color={rgb, 1: red, 0.06666666666666667;green, 0.4666666666666667;blue, 0.2}},, line width=\linewidthtikz}
\addlegendentry{\ifthenelse{\boolean{LegendAutomatic}}{Distributed JSCC}{\legendlabelb} }
 
\addlegendimage{color={rgb, 1: red, 0.6666666666666666;green, 0.26666666666666666;blue, 0.6}, mark size=\marksize, mark=square, mark options={ solid,color={rgb, 1: red, 0.6666666666666666;green, 0.26666666666666666;blue, 0.6}},, line width=\linewidthtikz}
\addlegendentry{\ifthenelse{\boolean{LegendAutomatic}}{Semantic JSCC}{\legendlabelc} }
 
\addlegendimage{color={rgb, 1: red, 0.2;green, 0.13333333333333333;blue, 0.5333333333333333}, mark size=\marksize, mark=asterisk, mark options={ solid,color={rgb, 1: red, 0.2;green, 0.13333333333333333;blue, 0.5333333333333333}},, line width=\linewidthtikz}
\addlegendentry{\ifthenelse{\boolean{LegendAutomatic}}{Semantic JSCC AirComp}{\legendlabeld} }
 
\addlegendimage{color={rgb, 1: red, 0.5333333333333333;green, 0.13333333333333333;blue, 0.3333333333333333}, mark size=\marksize, mark=Mercedes star flipped, mark options={ solid,color={rgb, 1: red, 0.5333333333333333;green, 0.13333333333333333;blue, 0.3333333333333333}},, line width=\linewidthtikz}
\addlegendentry{\ifthenelse{\boolean{LegendAutomatic}}{Error-free communication}{\legendlabele} }
 
\end{axis} 
\end{tikzpicture}
    \vspace{-0.3cm}
    
    \begin{subfigure}{0.49\columnwidth}
        \centering
        \caption{$N_\text{ch}=91$}
        \includegraphics[width=1.0\columnwidth]{tikz/atc_fwi_eval/atc_fwi_over_snr_91fc.tikz}
        \label{fig:over_snr_91fc}
    \end{subfigure}
    \hfill
    \begin{subfigure}{0.49\columnwidth}
        \centering
        \caption{$N_\text{ch}=45$}
        \includegraphics[width=1.0\columnwidth]{tikz/atc_fwi_eval/atc_fwi_over_snr_45fc.tikz}
        \label{fig:over_snr_45fc}
    \end{subfigure}

    \vspace{-0.3cm}

   %  \begin{subfigure}{0.49\columnwidth}
   %      \centering
   %      \caption{Cost}
			% \includegraphics[width=1.0\columnwidth]{tikz/atc_fwi_eval/atc_fwi_over_snr_45.tikz}
   %  \end{subfigure}
   %  \hfill
   %  \begin{subfigure}{0.49\columnwidth}
   %      \centering
   %      \caption{Cost}
			% \includegraphics[width=1.0\columnwidth]{tikz/atc_fwi_eval/atc_fwi_over_snr_45fc.tikz}
   %  \end{subfigure}

    \caption{\ac{ATC-FWI} algorithm performance: 
    NMSE of the estimated velocity averaged over all agents and the \ac{ATC-FWI} cost function after $35$ algorithm iterations over the communication channel SNR is shown for a MAS of eight agents with a full mesh communication topology (seven communicating neighbors for each agent).}
    \label{fig:atc_fwi_over_SNR}
\end{figure}

\subsubsection{Exemplary result of ATC-FWI for block anomaly}
Finally, we show the estimated velocity models for a model with a squared anomaly of $\SI{3.20}{\kilo \meter / \second}$  over a background velocity of $\SI{1.45}{\kilo \meter / \second}$. The results shown in Fig. \ref{fig:Block_anomaly} are after $40$ iterations of the \ac{ATC-FWI} algorithm. For all compared methods the estimated velocity models are shown in absolute terms. Furthermore, we depict the difference of each communication method in the imaging results when using error-free communication. The results demonstrate that when errors occur during communication, the velocity of the anomaly is underestimated on average. 
We observe that our proposed semantic \ac{JSCC} \ac{OAC} outperforms all other methods. In particular, the standard digital communication system fails completely.

\begin{figure}
    \centering
    \begin{subfigure}{\linewidth}
        \centering
        \caption{Imaging results}
        \begin{adjustbox}{right}
            \includegraphics[width=\linewidth]{tikz/png_images/block_anomaly_difference_to_perfect_comm_a.tikz}
        \end{adjustbox}
        \label{fig:Block_anomaly_a}
    \end{subfigure}

    \vspace{-0.0cm}

    \begin{subfigure}{\linewidth}
        \centering
        \caption{Imaging difference to the case of error-free communication}
        \begin{adjustbox}{right}
            \includegraphics[width=\linewidth]{tikz/png_images/block_anomaly_difference_to_perfect_comm_b.tikz}
        \end{adjustbox}
        \label{fig:Block_anomaly_b}
    \end{subfigure}
    \caption{
    In (a), the imaging result of the \ac{ATC-FWI} algorithm for the case of a block anomaly (ground truth, top left image) is shown. 
    In (b), the imaging difference to the error-free communication imaging result is shown.
    A \ac{MAS} of eight agents is used with a full mesh communication topology with $N_\text{ch}=91$.
    The results are shown for the agent positioned at the red arrow after $40$ iterations of the \ac{ATC-FWI} algorithm.
    }
    \label{fig:Block_anomaly}
\end{figure}

\section{Conclusion}

In this work, we proposed a semantic communication system for distributed function computation within the \ac{ATC-FWI}, a method that allows cooperative subsurface imaging within a \ac{MAS}. By means of numerical simulations, we showed that our proposed semantic \ac{JSCC} \ac{OAC} can outperform both classical \ac{SSCC} digital communication systems and state-of-the-art \ac{JSCC} methods.
For the imaging performance of the distributed \ac{ATC-FWI} algorithm, we showed that our proposed semantic methods using side information significantly outperform the generic \ac{JSCC} communication approaches for a strict compression rate. The classic \ac{SSCC} digital communication method even failed in the imaging task for the strict compression rates.
For the case of seven transmitting neighboring agents it was demonstrated that the distributed \ac{JSCC} approach could be outperformed by the proposed semantic \ac{JSCC} approach, which again was outperformed by the proposed semantic \ac{JSCC} \ac{OAC} approach.
We note that the proposed approach scales to larger \acp{MAS}, since parameter sharing is employed for the encoders and decoders during training. This prevents the number of trainable parameters from growing with the number of agents.
Finally, we mention again, that the investigation on how much a synchronization offset would limit the performance of the proposed semantic \ac{JSCC} \ac{OAC} approach is left as future work.

% \begin{itemize}
%     \item Unfortunately, as the correlation between the velocity models is mostly positive, the advantage of the semantic \ac{JSCC} approach to the distributed \ac{JSCC} approach is not significant \cite{Distributed_compression_gap_to_Optimal_Sum_Rate}. However, 

%     \item Significant performance gain using side information semantic function computation for averaging velocity models at decoder/receiver. Higher performance gain for velocity models due to higher correlation, less visible for gradients (due to lower correlation)

%     \item Imaging performance of ATC-FWI improved slightly with semantic approaches, more significant improvement with AirComp compared to distributed JSCC. Significant enhancement compared to JSCC without side info
    
%     \item Semantic \ac{JSCC} with \ac{OAC} brings a significant advantage if the number of agents that communicate directly with each other is large. (for full mesh, large no. of neighboring agents, higher network connectivity) However, a synchronization is required.

% \end{itemize}

% or


%\newpage
\appendices
\section{Derivation of lower bound of conditional mutual information}
%\appendix[Derivation of lower bound of conditional mutual information]
%\renewcommand{\theequation}{A.\arabic{equation}}
%\setcounter{equation}{0}
For the following derivation, we use $\bm{z}$ for $\gradSum$ and $\bm{s}_0$ for $\gradIndexZero$ in case of the velocity gradients, and $\bm{z}$ for $\veloSum$ and $\bm{s}_0$ for $\veloIndexZero$, in case of the velocity models.
\label{Appendix:derivationMILowerbound}
\begin{samepage}
\begin{align}
 I(\bm{z};\bm{y} |\bm{\s}_0)
                            & =  E_{p(\bm{z},\bm{y} ,\bm{\s}_0)}\left[ \log\left( \frac{p(\bm{z}|\bm{y} ,\bm{\s}_0)}{p(\bm{z}|\bm{\s}_0)} \right) \right] \\
                            & = E_{p(\bm{z},\bm{y} ,\bm{\s}_0)}\left[ \log\left( \frac{p(\bm{z}|\bm{y} ,\bm{\s}_0)}{p(\bm{z}|\bm{\s}_0)} \frac{q_{\bm{\phi}}(\bm{z}|\bm{y} ,\bm{\s}_0)}{q_{\bm{\phi}}(\bm{z}|\bm{y} ,\bm{\s}_0)}\right) \right]\\
                            &= E_{p(\bm{z},\bm{y} ,\bm{\s}_0)}\left[ \log\left( q_{\bm{\phi}}(\bm{z}|\bm{y} ,\bm{\s}_0)\right) \right] \\
                            & \quad \quad - E_{p(\bm{z},\bm{y} ,\bm{\s}_0)}\left[ \log\left( p(\bm{z}|\bm{\s}_0)\right) \right] \nonumber \\
                            & \quad \quad + E_{p(\bm{z},\bm{y} ,\bm{\s}_0)}\left[ \log\left( \frac{p(\bm{z}|\bm{y} ,\bm{\s}_0) }{q_{\bm{\phi}}(\bm{z}|\bm{y} ,\bm{\s}_0)}\right) \right] \nonumber\\
                            & = E_{p(\bm{y} ,\bm{\s}_0)}\left[ E_{p(\bm{z}|\bm{y} ,\bm{\s}_0)}\left[ \log\left( q_{\bm{\phi}}(\bm{z}|\bm{y} ,\bm{\s}_0)\right) \right] \right] \\
                            & \quad \quad - E_{p(\bm{z},\bm{\s}_0)}\left[ \log\left( p(\bm{z}|\bm{\s}_0)\right) \right] \nonumber \\
                            & \quad \quad + E_{p(\bm{z},\bm{y} ,\bm{\s}_0)}\left[ \log\left( \frac{p(\bm{z}|\bm{y} ,\bm{\s}_0) }{q_{\bm{\phi}}(\bm{z}|\bm{y} ,\bm{\s}_0)}\right) \right] \nonumber\\
                            & = - E_{p(\bm{y} ,\bm{\s}_0)}\left[H \left( p(\bm{z}|\bm{y} ,\bm{\s}_0), q_{\bm{\phi}}(\bm{z}|\bm{y} ,\bm{\s}_0)\right) \right] \label{condMI_identity} \\
                            & \quad \quad +  E_{p(\bm{\s}_0)}\left[ H(p(\bm{z}|\bm{\s}_0)) \right] \nonumber \\
                            & \quad \quad + E_{p(\bm{y},\bm{\s}_0)}\left[D_{\text{KL}}\left( p(\bm{z}|\bm{y} ,\bm{\s}_0) || q_{\bm{\phi}}(\bm{z}|\bm{y} ,\bm{\s}_0) \right)   \right]\nonumber\\
                            &\geq - E_{p(\bm{y} ,\bm{\s}_0)}\left[H \left( p(\bm{z}|\bm{y} ,\bm{\s}_0), q_{\bm{\phi}}(\bm{z}|\bm{y} ,\bm{\s}_0)\right)\right]  \label{MILBo} \\
                            & \quad \quad + E_{p(\bm{\s}_0)}\left[H(p(\bm{z}|\bm{\s}_0)) \right], \nonumber
\end{align}
\end{samepage}
where $E_{p(x)}[\cdot]$ is the expected value with respect to $x\sim p(x)$, $H(\cdot)$ is the Shannon entropy, $H(q,p)$ is the cross entropy between $p$ and $q$, and $D_{\text{KL}}(p||q)$ is the \ac{KL} divergence from $p$ to $q$.
The step for \eqref{MILBo} uses the fact that the \ac{KL} divergence is non-negative \cite{cover1999informationTheory}. 

% Dont Delte: extra derivation that can be omited in paper
% Using equation \eqref{condMI_identity} and
% \begin{equation}
%     E_{p(\bm{z},\bm{y} ,\bm{\s}_0)}\left[ \log\left( \frac{p(\bm{z}|\bm{y} ,\bm{\s}_0)}{p(\bm{z}|\bm{\s}_0)} \right) \right] = H(\bm{z}|\bm{\s}_0) - H(\bm{z}|\bm{y},\bm{\s}_0),
% \end{equation}

\section{Neural Network Training}
\label{Appendix:nn_training}
% We also mention again that in all digital computer systems all calculations are done in the discrete domain, e.g. with float $32$ values. However, float $32$ values give about $2^{32 \cdot5000}$ different events. Together with a vector size of $5000$ from a grid of $100$ by $50$ a discrete space of about 
%  $2^{32 \cdot5000}$ different events occurs, which are simply too many parameters to estimate with a neural network. To make the discrete probability estimation feasible, one would need to assume that each output is indepdendent of the output at the other grid points and that the discretization uses way less discretization levels. However, even if independence is assumed as an $8$ bit quantization level, still $5000\cdot 2^8 = 1.28 \cdot 10^{6}$ parameters have to be estimated, which would still require enormous computational effort. 

%\begin{itemize}
%    \item Reparameteraization Trick (to do)
%    \item solve with gradient approximated by batchse in NN training (to do)
%    \item assume Gaussian with mean as parametesr output from DNN leads to MSE loss (is there)
%\end{itemize}

\label{Implementation_considerations}
%In this section, we show how the gradients with respect to the encoder and decoder parameters can be estimated efficiently from the training data. To train the \ac{NN}, training data samples of the observations $\bm{\s}_0, \veloIndexOne,\dots,\bm{\s}_{\numberReceivers}$ are used as the inputs, and training data samples of the semantic variable $\bm{z}$ are used as the desired output. Furthermore, a suitable type of parameterized distribution for the decoder is selected such that \ac{NN} training is computationally feasible.

In this section, we show how the gradients with respect to the encoder and decoder parameters can be estimated efficiently from the training data. 
Here, we use $\bm{z}$ for $\gradSum$ and $\bm{s}_0, \bm{\s}_1,\dots,\bm{\s}_{\numberReceivers}$ for $\gradIndexZero, \gradIndexOne, \dots, \gradIndexLast$ in case of the velocity gradients, and $\bm{z}$ for $\veloSum$ and $\bm{s}_0, \bm{\s}_1,\dots,\bm{\s}_{\numberReceivers}$ for $\veloIndexZero, \veloIndexOne, \dots, \veloIndexLast$, in case of the velocity models.
%For the velocity gradients, $\bm{z}$ is $\gradSum$ and $\bm{s}_0, \bm{\s}_1,\dots,\bm{\s}_{\numberReceivers}$ are $\gradIndexZero, \gradIndexOne, \dots, \gradIndexLast$, and for the velocity models, $\bm{z}$ is $\veloSum$ and $\bm{s}_0, \bm{\s}_1,\dots,\bm{\s}_{\numberReceivers}$ are $\veloIndexZero, \veloIndexOne, \dots, \veloIndexLast$. 
To train the \ac{NN}, training data samples of the observations $\bm{\s}_0, \bm{\s}_1,\dots,\bm{\s}_{\numberReceivers}$ are used as the inputs, and training data samples of the semantic variable $\bm{z}$ are used as the desired output.

Using \ac{SGD} based optimization of the autoencoder, the gradient with respect to the decoder parameters ${\bm{\phi}}$ can be approximated for a batch of training data by
\begin{align}
    & \quad \frac{ \partial}{\partial {\bm{\phi}}}  E_{p(\bm{y} ,\bm{\s}_0)}\left[H\left( p(\bm{z}|\bm{y} ,\bm{\s}_0), q_{\bm{\phi}}(\bm{z}|\bm{y} ,\bm{\s}_0)\right)\right] \\
    %& = -\frac{ \partial}{\partial {\bm{\phi}}} E_{p(\bm{y} ,\bm{\s}_0)}\left[ E_{p(\bm{z}|\bm{y} ,\bm{\s}_0)}\left[ \log\left( q_{\bm{\phi}}(\bm{z}|\bm{y} ,\bm{\s}_0)\right) \right] \right] \\
    %& = -\frac{ \partial}{\partial {\bm{\phi}}} E_{p(\bm{z},\bm{y} ,\bm{\s}_0)}\left[ \log\left( q_{\bm{\phi}}(\bm{z}|\bm{y} ,\bm{\s}_0)\right) \right] \\
    %& \approx -\sum_{k=1}^K \frac{\partial}{\partial {\bm{\phi}}} E_{p(\bm{z}|\bm{y} ,\bm{\s}_0)}\left[ \log\left( q_{\bm{\phi}}(\bm{z}|\bm{y} ,\bm{\s}_0)\right) \right] \\
    & = - E_{p(\bm{z},\bm{y} ,\bm{\s}_0)}\left[ \frac{ \partial \log\left( q_{\bm{\phi}}(\bm{z}|\bm{y} ,\bm{\s}_0)\right)}{\partial {\bm{\phi}}}  \right]  \\
    & \approx -\frac{1}{L}\sum_{l=1}^L \frac{ \partial \log\left( q_{\bm{\phi}}(\bm{z}_l|\bm{y}_l ,\bm{\s}_{0_l})\right)}{\partial {\bm{\phi}}}, \label{sample_average_loss_derivative}
\end{align}
where $L$ is the batch size and $\bm{z}_l$, $\bm{y}_l$, and $\bm{\s}_{0_l}$ are training data samples from the distribution $p(\bm{z},\bm{y},\bm{\s}_0)$.

To estimate the gradients with respect to the encoders with parameters ${\bm{\theta}_1},\dots,  {\bm{\theta}_{\numberReceivers}}$, we sample from $p_{{\bm{\theta}_\idxNeigh}}(\bm{c}_\idxNeigh|\bm{\s}_\idxNeigh)$, which depends itself on ${\bm{\theta}_\idxNeigh}$. This leads to high variance in the gradient estimation \cite{simeone2022machine}.

%When calculating the gradients with respect to the encoders with parameters ${\bm{\theta}_1},\dots,  {\bm{\theta}_{\numberReceivers}}$, the derivative with respect to ${\bm{\theta}_\idxNeigh}$ cannot be put into the expectation, as $\bm{y}$ depends on ${\bm{\theta}_1},\dots,  {\bm{\theta}_{\numberReceivers}}$, as $\bm{c}_\idxNeigh \sim p_{{\bm{\theta}_\idxNeigh}}(\bm{c}_\idxNeigh|\bm{\s}_\idxNeigh)$ and $\bm{y} \sim p(\bm{y}|\bm{c}_1, \dots , \bm{c}_{\numberReceivers})$. The gradients can still be estimated in that case, but this leads to the problem of high variance of the gradient estimates \cite{simeone2022machine}. 
To get less noisy gradient estimates for the \ac{SGD} optimizer, the reparametrization trick can be used by separating $\bm{y} \sim p(\bm{y}|\bm{c}_1,\dots,\bm{c}_{\numberReceivers})$ into a deterministic and differentiable function $f_{\bm{\theta}}(\bm{\s},\bm{n})$ with random variable $\bm{n}$, which is independent of the $\bm{\theta}_\idxNeigh$, and $\bm{\theta}=\begin{bmatrix}
    \bm{\theta}_1^\top,\dots,\bm{\theta}_{\numberReceivers}^\top
\end{bmatrix}^\top$, $\bm{s} = \begin{bmatrix}
    \bm{s}_1^\top,\dots,\bm{s}_{\numberReceivers}^\top
\end{bmatrix}^\top$ \cite{simeone2022machine}.
In our case, we can apply the reparametrization trick by assuming deterministic encoders $p_{\bm{\theta}_\idxNeigh}(\bm{c}_\idxNeigh|\bm{s}_\idxNeigh)$ and additive noise $\bm{n}$ for the wireless channel. 
Then the gradients with respect to ${\bm{\theta}_\idxNeigh}$ can be written as
%For the reparametrization trick, $\bm{y}$ can be written as a sum of a deterministic function $f_\theta$ with parameters ${\bm{\theta}_1},\dots,  {\bm{\theta}_{\numberReceivers}}$ and a Gaussian distributed random variable $\bm{n}$, which lets us write $\bm{y}=f_\theta(\bm{\s}_1,\dots,\bm{\s}_{\numberReceivers}) + \bm{n}$.
%This way the expected value is with respect to the random variable $\bm{n}$ instead of $\bm{y}$, which gives the gradient with respect to ${\bm{\theta}_\idxNeigh}$ as
\begin{align}
    & \quad \frac{ \partial}{\partial {\bm{\theta}_\idxNeigh}}  E_{p(\bm{y} ,\bm{\s}_0)}\left[H\left( p(\bm{z}|\bm{y} ,\bm{\s}_0), q_{\bm{\phi}}(\bm{z}|\bm{y} ,\bm{\s}_0)\right)\right] \\
    %& = \!-\frac{ \partial}{\partial {\bm{\theta}_\idxNeigh}} E_{p(\bm{z},\bm{\s})p(\bm{n})}\!\left[ \log\left[ q_{\bm{\phi}}(\bm{z}|f_{\bm{\theta}}(\bm{\s},\bm{n}),\bm{\s}_0)\right] \right] \nonumber  \\
    & \!= - E_{p(\bm{z},\bm{\s}_0,\bm{\s}_1,\dots,\bm{\s}_{\numberReceivers})p(\bm{n})}\! \!\left[ \frac{\partial f_{\bm{\theta}}(\bm{\s},\bm{n})}{\partial \bm{\theta}_\idxNeigh} \frac{ \partial \log\left[ q_{\bm{\phi}}(\bm{z}|\bm{y} ,\bm{\s}_0)\right]}{\partial {\bm{y}}}  \right]  \\
    & \approx -\frac{1}{L}\sum_{l=1}^L \frac{\partial f_{\bm{\theta}}(\bm{\s}_l,\bm{n}_l)}{\partial \bm{\theta}_\idxNeigh} \left. \frac{ \partial \log\left[ q_{\bm{\phi}}(\bm{z}_l|\bm{y} ,\bm{\s}_{0_l})\right]}{\partial {\bm{y}}} \right|_{\bm{y}=f_{\bm{\theta}}(\bm{\s}_l,\bm{n}_l)}, \label{reparametrization_trick_applied}
\end{align}
where $\bm{\s}_l$ and $\bm{n}_l$ are training data samples of $\bm{\s}$ and $\bm{n}$, respectively.
%where $K$ is the batch size and $\bm{z}_l$, $\bm{y}_l$, and $\bm{\s}_{0_l}$ are samples from the distribution $p(\bm{z},\bm{y},\bm{\s}_0)$.

\section{Derivation of the optimization objective under Gaussian assumption}
\label{Derivation_CE_under_Gaussian}
Here, we use $\bm{z}$ for $\gradSum$ and $\bm{s}_0$ for $\gradIndexZero$ in case of the velocity gradients, and $\bm{z}$ for $\veloSum$ and $\bm{s}_0$ for $\veloIndexZero$, in case of the velocity models.
To be able to optimize our communication system, a simplification for the decoder probability density function $q_{\bm{\phi}}(\bm{z}|\bm{y},\bm{\s}_0)$ is required. Here, we assume independent Gaussian distributions for every $i$-th element $z_i$ of $\bm{z}\in\R^{\Nx\Nz}$ for $q_{\bm{\phi}}(\bm{z}|\bm{y} ,\bm{\s}_{0})$ with fixed variance $\sigma^2$ and means $\bm{\mu}(\bm{y},\bm{\s}_0)$.
The $i$-th mean $\mu_{i}(\bm{y},\bm{\s}_0)$ is the $i$-th output of the decoder $\bm{\mu}(\bm{y},\bm{\s}_0)$, which depends on the decoder inputs $\bm{y}$ and $\bm{\s}_0$. 
Given this assumption, our optimization objective of the cross entropy can be simplified to
\begin{align}
    & E_{p(\bm{y} ,\bm{\s}_0)}\left[H\left( p(\bm{z}|\bm{y} ,\bm{\s}_0), q_{\bm{\phi}}(\bm{z}|\bm{y} ,\bm{\s}_0)\right)\right] \\
    = &E_{p(\bm{z},\bm{y} ,\bm{\s}_0)}\left[  \log\left( q_{\bm{\phi}}(\bm{z}|\bm{y} ,\bm{\s}_{0})\right) \right]\\
     = &E_{p(\bm{z},\bm{y} ,\bm{\s}_0)}\left[  \log \!\left( \prod_{i=1}^{\Nx\Nz}  \frac{1}{\sqrt{2\pi\sigma^2}} \exp \! \left( -\frac{\left(z_i - \mu_{i}(\bm{y},\bm{\s}_0)\right)^2}{2\sigma^2} \! \right) \! \right) \!\right]\\
     = &E_{p(\bm{z},\bm{y} ,\bm{\s}_0)} \!\!\left[ \! \frac{1}{2\sigma^2}\sum_{i=1}^{\Nx\Nz} \!\left(z_i - \mu_{i}(\bm{y},\bm{\s}_0)\right)^2 \! \right] \!\! + \! \Nx\Nz\log \!\sqrt{2\pi\sigma^2} \\
     =& \frac{1}{2\sigma^2} E_{p(\bm{z},\bm{y} ,\bm{\s}_0)}\left[ \|\bm{z} - \bm{\mu}(\bm{y},\bm{\s}_0) \|_2^2 \right] + \Nx\Nz\log\sqrt{2\pi\sigma^2}.\label{objective_MSE_appendix}
\end{align}


%\section*{Acknowledgment}


\ifCLASSOPTIONcaptionsoff
  \newpage
\fi



%\bibliographystyle{IEEEtran}
\bibliography{references}

\begin{thebibliography}{10}
\providecommand{\url}[1]{#1}
\csname url@samestyle\endcsname
\providecommand{\newblock}{\relax}
\providecommand{\bibinfo}[2]{#2}
\providecommand{\BIBentrySTDinterwordspacing}{\spaceskip=0pt\relax}
\providecommand{\BIBentryALTinterwordstretchfactor}{4}
\providecommand{\BIBentryALTinterwordspacing}{\spaceskip=\fontdimen2\font plus
\BIBentryALTinterwordstretchfactor\fontdimen3\font minus
  \fontdimen4\font\relax}
\providecommand{\BIBforeignlanguage}[2]{{%
\expandafter\ifx\csname l@#1\endcsname\relax
\typeout{** WARNING: IEEEtran.bst: No hyphenation pattern has been}%
\typeout{** loaded for the language `#1'. Using the pattern for}%
\typeout{** the default language instead.}%
\else
\language=\csname l@#1\endcsname
\fi
#2}}
\providecommand{\BIBdecl}{\relax}
\BIBdecl

\bibitem{Zhang2020}
\BIBentryALTinterwordspacing
S.~Zhang, R.~Pohlmann, T.~Wiedemann, A.~Dammann, H.~Wymeersch, and P.~A.
  Hoeher, ``Self-aware swarm navigation in autonomous exploration missions,''
  \emph{Proceedings of the IEEE}, vol. 108, no.~7, p. 1168–1195, Jul. 2020.
  [Online]. Available: \url{http://dx.doi.org/10.1109/JPROC.2020.2985950}
\BIBentrySTDinterwordspacing

\bibitem{Nathan2024}
\BIBentryALTinterwordspacing
R.~J. A.~A. Nathan, S.~Strand, D.~Mehrwald, D.~Shutin, and O.~Bimber, ``An
  autonomous drone swarm for detecting and tracking anomalies among dense
  vegetation,'' 2024. [Online]. Available:
  \url{https://arxiv.org/abs/2407.10754}
\BIBentrySTDinterwordspacing

\bibitem{Rabideau2025}
\BIBentryALTinterwordspacing
G.~Rabideau, J.~Russino, A.~Branch, N.~Dhamani \emph{et~al.}, ``Planning,
  scheduling, and execution on the moon: the cadre technology demonstration
  mission,'' 2025. [Online]. Available: \url{https://arxiv.org/abs/2502.14803}
\BIBentrySTDinterwordspacing

\bibitem{Wiedemann2019}
\BIBentryALTinterwordspacing
T.~Wiedemann, D.~Shutin, and A.~J. Lilienthal, ``Model-based gas source
  localization strategy for a cooperative multi-robot system—a probabilistic
  approach and experimental validation incorporating physical knowledge and
  model uncertainties,'' \emph{Robotics and Autonomous Systems}, vol. 118, p.
  66–79, Aug. 2019. [Online]. Available:
  \url{http://dx.doi.org/10.1016/j.robot.2019.03.014}
\BIBentrySTDinterwordspacing

\bibitem{Shin2024}
\BIBentryALTinterwordspacing
B.-S. Shin and D.~Shutin, ``Joint distributed traveltime and full waveform
  tomography for enhanced subsurface imaging in seismic networks,'' \emph{IEEE
  Transactions on Computational Imaging}, vol.~10, p. 600–612, 2024.
  [Online]. Available: \url{http://dx.doi.org/10.1109/TCI.2024.3366754}
\BIBentrySTDinterwordspacing

\bibitem{Rizk2025}
\BIBentryALTinterwordspacing
E.~Rizk, K.~Yuan, and A.~H. Sayed, ``Diffusion learning with partial agent
  participation and local updates,'' 2025. [Online]. Available:
  \url{https://arxiv.org/abs/2505.11307}
\BIBentrySTDinterwordspacing

\bibitem{Zhang2025}
\BIBentryALTinterwordspacing
J.~Zhang, Z.~Liu, Y.~Zhu, E.~Shi, B.~Xu, C.~Yuen, D.~Niyato, M.~Debbah, S.~Jin,
  B.~Ai, and S.~Xuemin, ``Multi-agent reinforcement learning in wireless
  distributed networks for 6g,'' 2025. [Online]. Available:
  \url{https://arxiv.org/abs/2502.05812}
\BIBentrySTDinterwordspacing

\bibitem{survey_gunduz2022}
D.~G{\"u}nd{\"u}z, Z.~Qin, I.~E. Aguerri, H.~S. Dhillon, Z.~Yang, A.~Yener,
  K.~K. Wong, and C.-B. Chae, ``Beyond transmitting bits: Context, semantics,
  and task-oriented communications,'' \emph{IEEE Journal on Selected Areas in
  Communications}, vol.~41, no.~1, pp. 5--41, 2022.

\bibitem{survey_yang2022}
W.~Yang, H.~Du, Z.~Q. Liew, W.~Y.~B. Lim, Z.~Xiong, D.~Niyato, X.~Chi, X.~Shen,
  and C.~Miao, ``Semantic communications for future internet: Fundamentals,
  applications, and challenges,'' \emph{IEEE Communications Surveys \&
  Tutorials}, vol.~25, no.~1, pp. 213--250, 2022.

\bibitem{survey_wheeler2023}
D.~Wheeler and B.~Natarajan, ``Engineering semantic communication: A survey,''
  \emph{IEEE Access}, vol.~11, pp. 13\,965--13\,995, 2023.

\bibitem{xu2023edge}
W.~Xu, Z.~Yang, D.~W.~K. Ng, M.~Levorato, Y.~C. Eldar, and M.~Debbah, ``Edge
  learning for b5g networks with distributed signal processing: Semantic
  communication, edge computing, and wireless sensing,'' \emph{IEEE journal of
  selected topics in signal processing}, vol.~17, no.~1, pp. 9--39, 2023.

\bibitem{Kalita2021}
\BIBentryALTinterwordspacing
H.~Kalita, A.~Quintero, A.~Wissing, B.~Haugh, C.~Angie, G.~Nail, J.~Wilson,
  J.~Richards, J.~Landin, K.~Kukkala, M.~Vazquez, N.~Tan, Q.~Lamey, R.~Lu,
  R.~Peralta, V.~Vilvanathan, and J.~Thangavelautham, ``Evaluation of lunar
  pits and lava tubes for use as human habitats,'' in \emph{Earth and Space
  2021}.\hskip 1em plus 0.5em minus 0.4em\relax American Society of Civil
  Engineers, Apr. 2021, p. 944–957. [Online]. Available:
  \url{http://dx.doi.org/10.1061/9780784483374.086}
\BIBentrySTDinterwordspacing

\bibitem{delaCroix2024}
\BIBentryALTinterwordspacing
J.-P. de~la Croix, F.~Rossi, R.~Brockers, D.~Aguilar, K.~Albee, E.~Boroson,
  A.~Cauligi, J.~Delaune, R.~Hewitt, D.~Kogan, G.~Lim, B.~Morrell, Y.~Nakka,
  V.~Nguyen, P.~Proen\c{c}a, G.~Rabideau, J.~Russino, M.~S. da~Silva, G.~Zohar,
  and S.~Comandur, ``Multi-agent autonomy for space exploration on the cadre
  lunar technology demonstration,'' in \emph{2024 IEEE Aerospace
  Conference}.\hskip 1em plus 0.5em minus 0.4em\relax IEEE, Mar. 2024, p.
  1–14. [Online]. Available:
  \url{http://dx.doi.org/10.1109/AERO58975.2024.10521425}
\BIBentrySTDinterwordspacing

\bibitem{Sayed2013}
A.~H. Sayed \emph{et~al.}, ``{Diffusion strategies for adaptation and learning
  over networks},'' \emph{IEEE Signal Processing Magazine}, pp. 155--171, 2013.

\bibitem{Shi2015}
\BIBentryALTinterwordspacing
W.~Shi, Q.~Ling, G.~Wu, and W.~Yin, ``Extra: An exact first-order algorithm for
  decentralized consensus optimization,'' \emph{SIAM Journal on Optimization},
  vol.~25, no.~2, p. 944–966, Jan. 2015. [Online]. Available:
  \url{http://dx.doi.org/10.1137/14096668X}
\BIBentrySTDinterwordspacing

\bibitem{ma2011some}
N.~Ma and P.~Ishwar, ``Some results on distributed source coding for
  interactive function computation,'' \emph{IEEE Transactions on Information
  Theory}, vol.~57, no.~9, pp. 6180--6195, 2011.

\bibitem{abari2016over}
O.~Abari, H.~Rahul, and D.~Katabi, ``Over-the-air function computation in
  sensor networks,'' \emph{arXiv preprint arXiv:1612.02307}, 2016.

\bibitem{vincent2010stacked}
P.~Vincent, H.~Larochelle, I.~Lajoie, Y.~Bengio, P.-A. Manzagol, and L.~Bottou,
  ``Stacked denoising autoencoders: Learning useful representations in a deep
  network with a local denoising criterion.'' \emph{Journal of machine learning
  research}, vol.~11, no.~12, 2010.

\bibitem{edgar2023semantic}
E.~Beck, C.~Bockelmann, and A.~Dekorsy, ``Semantic information recovery in
  wireless networks,'' \emph{Sensors}, vol.~23, no.~14, p. 6347, 2023.

\bibitem{Neural_DSC}
J.~Whang, A.~Nagle, A.~Acharya, H.~Kim, and A.~G. Dimakis, ``Neural distributed
  source coding,'' \emph{IEEE Journal on Selected Areas in Information Theory},
  vol.~5, pp. 493--508, 2024.

\bibitem{Virieux2009}
J.~Virieux and S.~Operto, ``{An overview of full-waveform inversion in
  exploration geophysics},'' \emph{Geophysics}, vol.~74, no.~6, 2009.

\bibitem{Fichtner2009}
A.~Fichtner, \emph{{Full Seismic Waveform Modelling and Inversion}}.\hskip 1em
  plus 0.5em minus 0.4em\relax Springer, Berlin, Heidelberg, 2009.

\bibitem{Li2020}
\BIBentryALTinterwordspacing
F.~Li, M.~Valero, Y.~Cheng, T.~Bai, and W.~Song, ``Distributed sensor networks
  based shallow subsurface imaging and infrastructure monitoring,'' \emph{IEEE
  Transactions on Signal and Information Processing over Networks}, vol.~6, p.
  241–250, 2020. [Online]. Available:
  \url{http://dx.doi.org/10.1109/TSIPN.2020.2975349}
\BIBentrySTDinterwordspacing

\bibitem{Wang2021}
\BIBentryALTinterwordspacing
S.~Wang, M.~P. Panning, S.~D. Vance, and W.~Song, ``Underground microseismic
  event monitoring and localization within sensor networks,'' \emph{Sensors},
  vol.~21, no.~8, p. 2830, Apr. 2021. [Online]. Available:
  \url{http://dx.doi.org/10.3390/s21082830}
\BIBentrySTDinterwordspacing

\bibitem{Plessix2006}
R.~E. Plessix, ``{A review of the adjoint-state method for computing the
  gradient of a functional with geophysical applications},'' \emph{Geophys. J.
  Int.}, 2006.

\bibitem{Louboutin2019}
\BIBentryALTinterwordspacing
M.~Louboutin, M.~Lange, F.~Luporini, N.~Kukreja, P.~A. Witte, F.~J. Herrmann,
  P.~Velesko, and G.~J. Gorman, ``Devito (v3.1.0): an embedded domain-specific
  language for finite differences and geophysical exploration,''
  \emph{Geoscientific Model Development}, vol.~12, no.~3, p. 1165–1187, Mar.
  2019. [Online]. Available: \url{http://dx.doi.org/10.5194/gmd-12-1165-2019}
\BIBentrySTDinterwordspacing

\bibitem{xu2023deep}
J.~Xu, T.-Y. Tung, B.~Ai, W.~Chen, Y.~Sun, and D.~G{\"u}nd{\"u}z, ``Deep joint
  source-channel coding for semantic communications,'' \emph{IEEE
  communications Magazine}, vol.~61, no.~11, pp. 42--48, 2023.

\bibitem{JSCC_survey}
D.~Gündüz, M.~A. Wigger, T.-Y. Tung, P.~Zhang, and Y.~Xiao, ``Joint
  source–channel coding: Fundamentals and recent progress in practical
  designs,'' \emph{Proceedings of the IEEE}, pp. 1--32, 2024.

\bibitem{razlighi2024cooperative}
A.~H. Razlighi, M.~H. Tillmann, E.~Beck, C.~Bockelmann, and A.~Dekorsy,
  ``Cooperative and collaborative multi-task semantic communication for
  distributed sources,'' \emph{arXiv preprint arXiv:2411.02150}, 2024.

\bibitem{Distributed_Image_Transmission}
S.~Wang, K.~Yang, J.~Dai, and K.~Niu, ``Distributed image transmission using
  deep joint source-channel coding,'' in \emph{ICASSP 2022 - 2022 IEEE
  International Conference on Acoustics, Speech and Signal Processing
  (ICASSP)}, 2022, pp. 5208--5212.

\bibitem{Mital_2023_WACV}
N.~Mital, E.~\"Ozyilkan, A.~Garjani, and D.~G\"und\"uz, ``Neural distributed
  image compression with cross-attention feature alignment,'' in
  \emph{Proceedings of the IEEE/CVF Winter Conference on Applications of
  Computer Vision (WACV)}, January 2023, pp. 2498--2507.

\bibitem{distrib_source_coding_FL}
H.~Yang, T.~Ding, and X.~Yuan, ``Federated learning with lossy distributed
  source coding: Analysis and optimization,'' \emph{IEEE Transactions on
  Communications}, vol.~71, no.~8, pp. 4561--4576, 2023.

\bibitem{FL_OAC}
K.~Yang, T.~Jiang, Y.~Shi, and Z.~Ding, ``Federated learning via over-the-air
  computation,'' \emph{IEEE Transactions on Wireless Communications}, vol.~19,
  no.~3, pp. 2022--2035, 2020.

\bibitem{OAC_channel_SNR}
O.~Abari, H.~Rahul, and D.~Katabi, ``Over-the-air function computation in
  sensor networks,'' \emph{arXiv preprint arXiv:1612.02307}, 2016.

\bibitem{Deep_OAC}
H.~Ye, G.~Y. Li, and B.-H.~F. Juang, ``Deep over-the-air computation,'' in
  \emph{GLOBECOM 2020 - 2020 IEEE Global Communications Conference}, 2020, pp.
  1--6.

\bibitem{OAC_semantic_multi_user}
S.~Wei, C.~Feng, C.~Guo, B.~Zhang, and J.~Chen, ``Deep over-the-air computation
  based multi-user semantic communication for smart cities,'' in \emph{2024
  IEEE/CIC International Conference on Communications in China (ICCC
  Workshops)}, 2024, pp. 622--626.

\bibitem{Distributed_semantic_oac_sensing}
P.~Yi, Y.~Cao, X.~Kang, and Y.-C. Liang, ``Integrated distributed semantic
  communication and over-the-air computation for cooperative spectrum
  sensing,'' \emph{IEEE Transactions on Communications}, pp. 1--1, 2024.

\bibitem{agrawal2024distributed}
N.~Agrawal, R.~L.~G. Cavalcante, M.~Yukawa, and S.~Sta{\'n}czak, ``Distributed
  convex optimization “over-the-air” in dynamic environments,'' \emph{IEEE
  Transactions on Signal and Information Processing over Networks}, vol.~10,
  pp. 610--625, 2024.

\bibitem{OAC_survey1}
Z.~Wang, Y.~Zhao, Y.~Zhou, Y.~Shi, C.~Jiang, and K.~B. Letaief, ``Over-the-air
  computation for 6g: Foundations, technologies, and applications,'' \emph{IEEE
  Internet of Things Journal}, vol.~11, no.~14, pp. 24\,634--24\,658, 2024.

\bibitem{OAC_survey2}
A.~Şahin and R.~Yang, ``A survey on over-the-air computation,'' \emph{IEEE
  Communications Surveys \& Tutorials}, vol.~25, no.~3, pp. 1877--1908, 2023.

\bibitem{cover1999informationTheory}
T.~M. Cover, \emph{Elements of information theory}.\hskip 1em plus 0.5em minus
  0.4em\relax John Wiley \& Sons, 1999.

\bibitem{simeone2022machine}
O.~Simeone, \emph{Machine learning for engineers}.\hskip 1em plus 0.5em minus
  0.4em\relax Cambridge university press, 2022.

\bibitem{poole2019boundsMI}
B.~Poole, S.~Ozair, A.~Van Den~Oord, A.~Alemi, and G.~Tucker, ``On variational
  bounds of mutual information,'' in \emph{International conference on machine
  learning}.\hskip 1em plus 0.5em minus 0.4em\relax PMLR, 2019, pp. 5171--5180.

\bibitem{barber2004algorithm}
D.~Barber and F.~Agakov, ``The im algorithm: a variational approach to
  information maximization,'' \emph{Advances in neural information processing
  systems}, vol.~16, no. 320, p. 201, 2004.

\bibitem{razlighi2024semantic}
A.~H. Razlighi, C.~Bockelmann, and A.~Dekorsy, ``Semantic communication for
  cooperative multi-task processing over wireless networks,'' \emph{IEEE
  Wireless Communications Letters}, 2024.

\bibitem{silverman2018density}
B.~W. Silverman, \emph{Density estimation for statistics and data
  analysis}.\hskip 1em plus 0.5em minus 0.4em\relax Routledge, 2018.

\bibitem{GoodfellowDeepLearning}
I.~Goodfellow, Y.~Bengio, and A.~Courville, \emph{Deep Learning}.\hskip 1em
  plus 0.5em minus 0.4em\relax MIT Press, 2016,
  \url{http://www.deeplearningbook.org}.

\bibitem{sionna}
J.~Hoydis, S.~Cammerer, F.~A. Aoudia, A.~Vem, N.~Binder, G.~Marcus, and
  A.~Keller, ``Sionna: An open-source library for next-generation physical
  layer research,'' \emph{arXiv preprint arXiv:2203.11854}, 2022.

\bibitem{jayaprakasam2017distributed}
S.~Jayaprakasam, S.~K.~A. Rahim, and C.~Y. Leow, ``Distributed and
  collaborative beamforming in wireless sensor networks: Classifications,
  trends, and research directions,'' \emph{IEEE Communications Surveys \&
  Tutorials}, vol.~19, no.~4, pp. 2092--2116, 2017.

\bibitem{li2021survey}
Z.~Li, F.~Liu, W.~Yang, S.~Peng, and J.~Zhou, ``A survey of convolutional
  neural networks: analysis, applications, and prospects,'' \emph{IEEE
  transactions on neural networks and learning systems}, vol.~33, no.~12, pp.
  6999--7019, 2021.

\bibitem{Gaussian_Two_Encoder_Source_Coding_Comparison}
A.~B. Wagner, S.~Tavildar, and P.~Viswanath, ``Rate region of the quadratic
  gaussian two-encoder source-coding problem,'' \emph{IEEE Transactions on
  Information Theory}, vol.~54, no.~5, pp. 1938--1961, 2008.

\bibitem{Distributed_compression_gap_to_Optimal_Sum_Rate}
Y.~Yang and Z.~Xiong, ``Distributed compression of linear functions: Partial
  sum-rate tightness and gap to optimal sum-rate,'' \emph{IEEE Transactions on
  Information Theory}, vol.~60, no.~5, pp. 2835--2855, 2014.

\bibitem{Shin2022}
B.-S. Shin and D.~Shutin, ``Distributed traveltime tomography using
  kernel-based regression in seismic networks,'' \emph{IEEE Geoscience and
  Remote Sensing Letters}, vol.~19, p. 1–5, 2022.

\end{thebibliography}
\bibliographystyle{IEEEtran}


\begin{IEEEbiography}[{\includegraphics[width=1in,height=1.25in,clip,keepaspectratio]{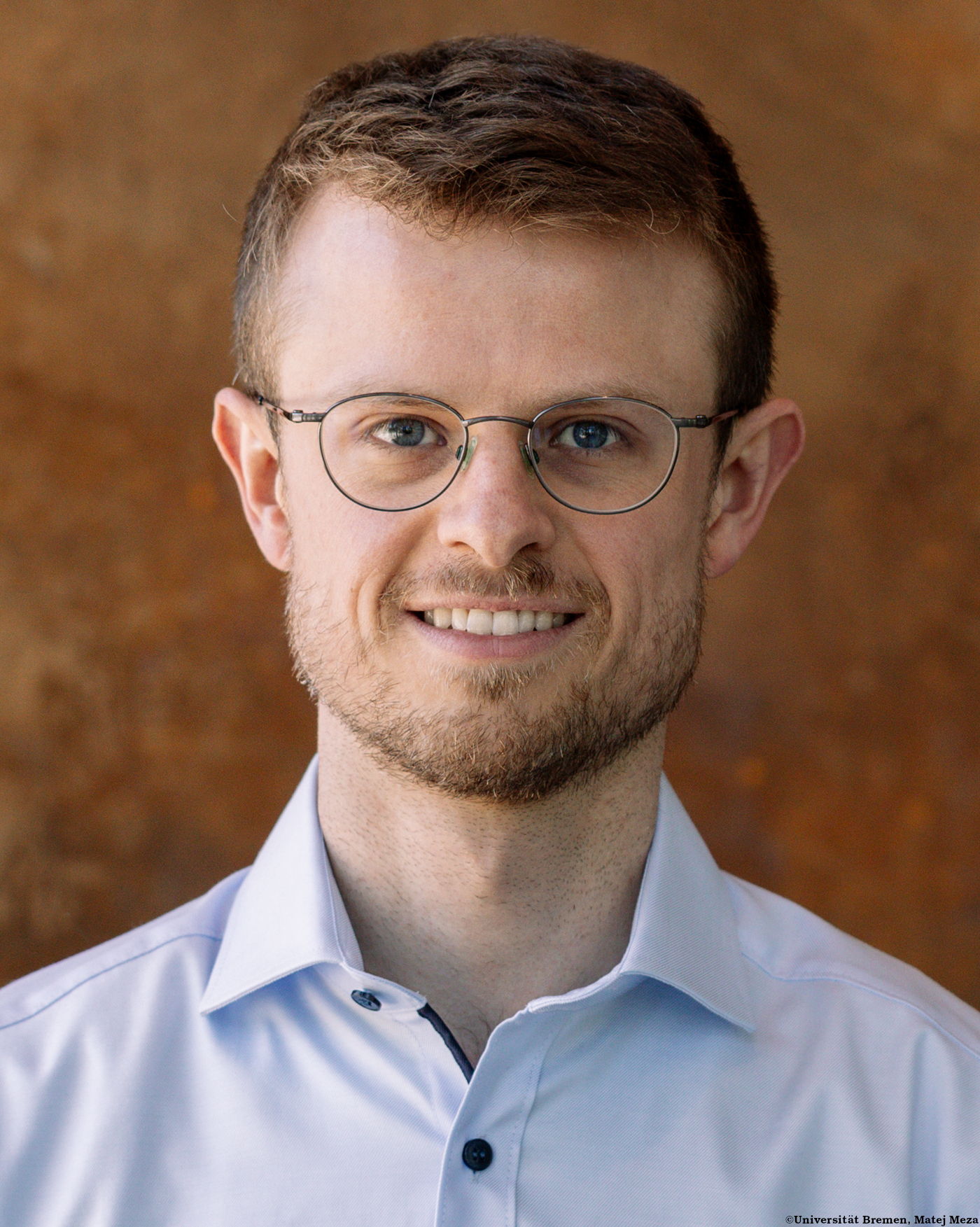}}]{Maximilian H. V. Tillmann}
(Graduate Student Member, IEEE) received the B.Sc. degree in elec-
trical engineering from RWTH Aachen University,
Aachen, Germany, in 2019, the M.E. degree from
Keio University, Yokohama, Japan, in 2022, with the
T.I.M.E double degree program, and the M.Sc. degree
from RWTH Aachen University in 2023. He is currently working towards the Ph.D. degree in electrical
engineering with the Department of Communications
Engineering (ANT), University of Bremen, Bremen,
Germany. His research interests include semantic communication, machine learning for wireless communication, and distributed optimization.

Mr. Tillmann was a recipient of the Friedrich Wilhelm Prize from RWTH Aachen University in 2024 for his master's thesis.
\end{IEEEbiography}

\begin{IEEEbiography}[{\includegraphics[width=1in,height=1.25in,clip,keepaspectratio]{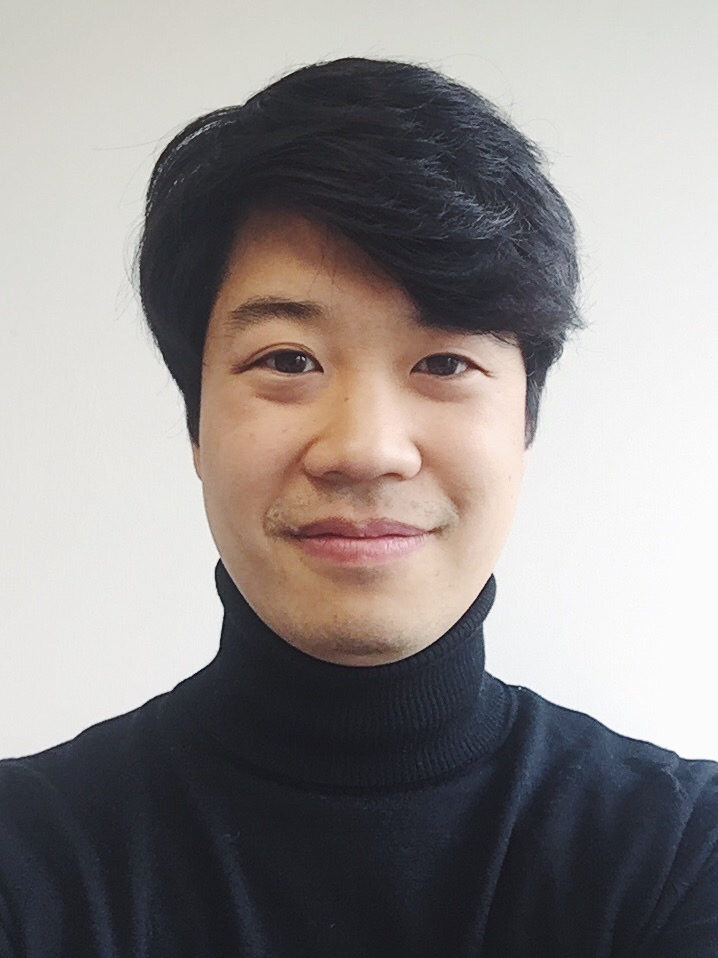}}]{Ban-Sok Shin}
(Member, IEEE) received the Dipl. Ing. (M.Sc.) and Dr.-Ing. (Ph.D.) degrees in electrical engineering from the University of Bremen, Bremen, Germany, in 2013 and 2020, respectively. From 2013 to 2019, he was a Research Assistant with the Department of Communications Engineering, University of Bremen, working on distributed estimation schemes for environmental monitoring applications. Since 2019, he has been a postdoctoral researcher in the Swarm Exploration Group at the Institute of Communications and Navigation, German Aerospace Center, where he is working on subsurface exploration and imaging using autonomous multi-agent networks. His research interests include subsurface imaging, in particular full waveform inversion, inverse problems, and distributed estimation, and their application to autonomous multi-agent networks.

Dr. Shin was the recipient of the OHB Award for the best Ph.D. dissertation from the Faculty of Electrical Engineering, University of Bremen, in 2021.
\end{IEEEbiography}

\begin{IEEEbiography}[{\includegraphics[width=1in,height=1.25in,clip,keepaspectratio]{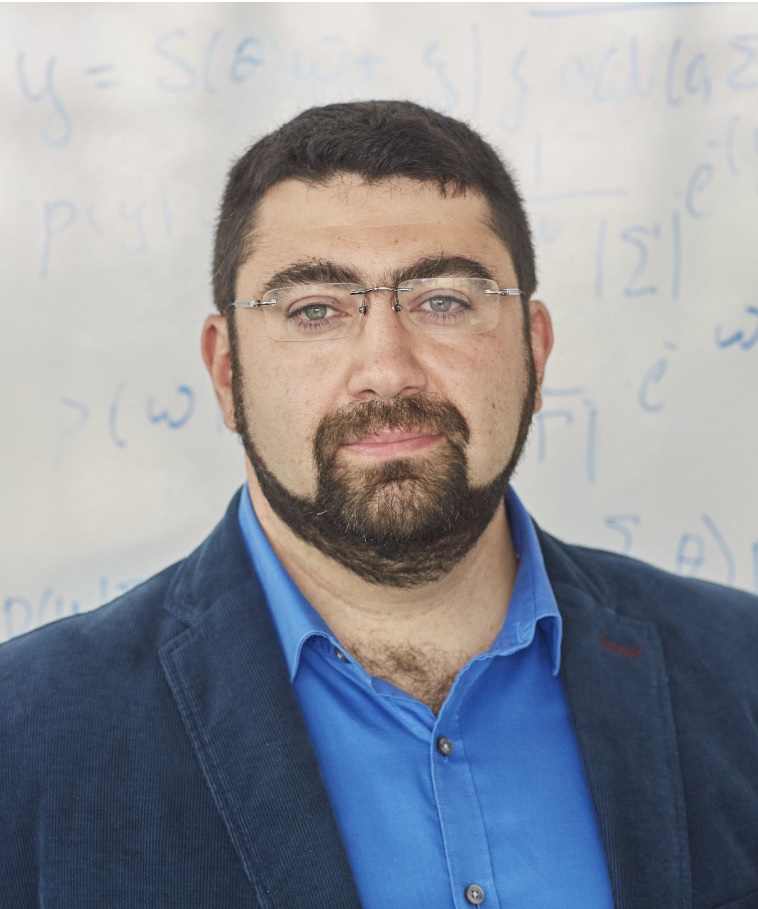}}]{Dmitriy Shutin}
(Senior Member, IEEE) received the master's degree in computer science from Dnipropetrovsk State University, Ukraine, in 2000, and the Ph.D. degree in electrical engineering from the Graz University of Technology, Graz, Austria, in 2006. From 2001 to 2006 and from 2006 to 2009, he was a Teaching Assistant and an Assistant Professor with the Signal Processing and Speech Communication Laboratory, Graz University of Technology. In 2009, he joined the Department of Electrical Engineering, Princeton University, where he worked as a Research Associate until 2011. In 2011, he joined the Institute of Communications and Navigation, German Aerospace Center, where he is currently a Leader of the Swarm Exploration Group. His research interests include machine learning for signal processing, distributed information and signal processing, and swarm exploration.

Dr. Shutin was a recipient of the Best Student Paper Award at the 2005 IEEE International Conference on Information, Communications and Signal Processing (ICICS). In 2009, he was awarded the Erwin Schroedinger Research Fellowship. From 2012 to 2014, he acted as a selected Advisor of German Air Navigation Service Provider within the Navigational System Panel of ICAO.
\end{IEEEbiography}

\begin{IEEEbiography}[{\includegraphics[width=1in,height=1.25in,clip,keepaspectratio]{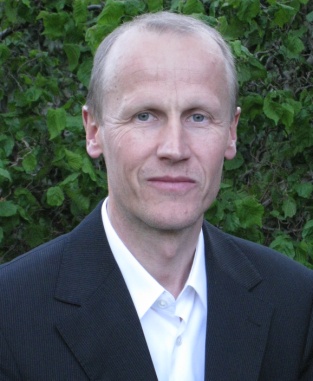}}]{Armin Dekorsy}
(Senior Member, IEEE) is a professor at the University of Bremen, where he is the director of the Gauss-Olbers Space Technology Transfer Center and heads the Department of Communications Engineering. With over eleven years of industry experience, including distinguished research positions such as DMTS at Bell Labs and Research Coordinator Europe at Qualcomm, he has actively participated in more than 65 international research projects, with leadership roles in 17 of them. He is a Senior Member of the IEEE Communications and Signal Processing Society and a member of the VDE/ITG Expert Committee on Information and System Theory. He co-authored the textbook \textit{Nachrichtenübertragung, Release 6, Springer Vieweg}, which is a bestseller in the field of communication technologies in German-speaking countries. His research focuses on signal processing and wireless communications for 5G/6G, industrial radio, and 3D networks.
\end{IEEEbiography}


% % insert where needed to balance the two columns on the last page with
% %\newpage



%\vfill


\end{document}